\documentclass[nofootinbib,twocolumn]{revtex4-1}

\usepackage{booktabs} 
\usepackage[top=1in, bottom=1in, left=1in, right=1in]{geometry} % Better use the page space. (1,1.25,1.25,1.25) is the standard margins.
\usepackage{graphicx}
\usepackage{color} % To use color
\usepackage{amsfonts,amssymb,amsmath,mathrsfs} % For additional math symbols
\usepackage{slashed,braket}       % For QM notation
\usepackage{bm}  % For bolded greek letters
\usepackage{bbm}  % For \mathbbm{1} 
\usepackage[colorlinks=true,urlcolor=blue
,anchorcolor=blue,citecolor=blue,filecolor=blue,linkcolor=blue,menucolor=blue
]{hyperref}
\usepackage[capitalize]{cleveref}
\usepackage{cancel}
\usepackage{fancyvrb} %for verbatim footnotes
\usepackage{xspace}
\usepackage{subfig}
\usepackage{ulem}
\usepackage[dvipsnames]{xcolor}
\usepackage{array}
\newcolumntype{L}{>{$}c<{$}}

\definecolor{c1}{rgb}{0., 0.26, 0.5}
\definecolor{c2}{rgb}{0.9,0.648, 0.}
\definecolor{c3}{rgb}{0.75, 0., 0.}
\definecolor{c4}{rgb}{0., 0.6, 0.6}
\definecolor{c5}{rgb}{0.357, 0.35, 0.7}
\definecolor{c6}{rgb}{0.588, 0.6, 0.}

\newcommand{\hc}{\mathrm{h.c.}}
\newcommand{\beq}{\begin{equation}}
\newcommand{\eeq}{\end{equation}}
\newcommand{\GeV}{\mathrm{GeV}}
\newcommand{\MeV}{\mathrm{MeV}}
\DeclareMathOperator\diag{diag}
\newcommand{\mulpsi}{\mu_{\ell\psi}}
\newcommand{\zmin}{z_{\mathrm{min}}}
\newcommand{\zmax}{z_{\mathrm{max}}}

\newcommand{\zdump}{z_{\mathrm{dump}}}

\newcommand{\gmtwo}{$(g-2)_\mu$\xspace}
\newcommand{\pythia}{\texttt{Pythia}\xspace}
\newcommand{\Npot}{N_p}

\begin{document}

\title{Diphoton Signals of Muon-philic Scalars at DarkQuest}

\author{Nikita Blinov}
\email{nblinov@yorku.ca} % 
\affiliation{Department of Physics and Astronomy, University of Victoria, Victoria, BC V8P 5C2, Canada}
\affiliation{Department of Physics and Astronomy, York University, Toronto, Ontario, M3J 1P3, Canada}
\author{Stefania Gori}
\email{sgori@ucsc.edu} % 

\author{Nick Hamer}
\email{nhamer@ucsc.edu} % 
\affiliation{Department of Physics, University of California Santa Cruz, 1156 High St., Santa Cruz, CA 95064, USA\\
and Santa Cruz Institute for Particle Physics, 1156 High St., Santa Cruz, CA 95064, USA
}

\date{\today}

%=============================================================================

%-----------------------------------------------------------------------------
\begin{abstract}
  We analyze the capability of the DarkQuest proton beam-dump experiment at Fermilab to discover new light resonances decaying into photons. As an example model, we focus on muon-philic scalar particles that decay to photons. This is one of the few minimal models that can address the $(g-2)_\mu$ anomaly at low mass. These scalars can be copiously produced by meson decays and muon bremsstrahlung. We point out that thanks to DarkQuest's compact geometry, muons can propagate through the dump and efficiently produce dark scalars near the end of the dump. This mechanism enables DarkQuest to be sensitive to both long-lived and prompt scalars. At the same time, di-photon signatures are generically not background free, and we discuss in detail the different sources of background and strategies to mitigate them. We find that the backgrounds can be sufficiently reduced for DarkQuest to test currently-viable $(g-2)_\mu$ parameter space.
\end{abstract}
%-----------------------------------------------------------------------------

\maketitle
\section{Introduction}
The existence of a dark sector, neutral under the Standard Model (SM) gauge symmetries, can offer an attractive explanation for Dark Matter (DM) and presents a compelling avenue for new physics exploration. This could be relevant for fundamental questions such as neutrino masses, the hierarchy problem, and the Universe's matter-antimatter asymmetry.

A vibrant set of efforts to look for sub-GeV dark sector particles - dark photons, scalars, and a variety of DM models - is now underway \cite{Gori:2022vri,Cooley:2022ufh,Coloma:2022dng}. An especially interesting direction of investigation has focused on the use of high-energy proton beams in fixed-target setups to produce and detect long-lived particles \cite{Gardner:2015wea,Berlin:2018pwi,Ahdida:2023okr}. In particular, it has been shown that the DarkQuest experiment, the upgrade of SpinQuest at Fermilab, will be able to probe a plethora of dark sector models \cite{Apyan:2022tsd}. So far, signatures with electrons and muons have been mainly investigated, with studies focusing on DM models, and on addressing the \gmtwo anomaly.

In fact, the measurement of the anomalous magnetic moment of the muon~\cite{Muong-2:2006rrc,Muong-2:2021ojo,Muong-2:2023cdq}, $a_\mu \equiv (g-2)_\mu/2$, remains 
in significant tension with Standard Model predictions. The difference between the Run I measurement  and the calculation from the \gmtwo theory initiative is~\cite{Aoyama:2020ynm}:
\begin{equation}
  a_\mu({\mathrm{Exp}}) -a_\mu({\mathrm{Theory}})=(251 \pm 59)\times 10^{-11}.
  \label{eq:g_minus_two_discrepancy}
\end{equation}
While recent lattice QCD results may point toward a potential problem with this SM prediction~\cite{Borsanyi:2020mff,Blum:2023qou,Bazavov:2023has}, new particles beyond the SM (BSM) still provide a compelling solution to this anomaly that is worth exploring further.

A large variety of new particle solutions have been proposed, including new vector bosons, and CP-even and CP-odd scalar particles.   
Because their effects on \gmtwo are through loops, this anomaly does not point to a very specific new physics scale and the viable mass range is very broad, spanning from $\sim$ MeV to several TeV~\cite{Capdevilla:2021kcf}. 
Therefore many experiments running at different energies are required to comprehensively probe this parameter space. 

The minimal requirement for a new particle to alleviate the \gmtwo problem is a non-negligible interaction with muons. Realistic 
models, however often contain couplings to other leptons, quarks or photons. These other interactions can provide a powerful way of 
testing these models; for example the minimal dark photon $(A')$ explanation of \gmtwo has been excluded 
in both visible~\cite{NA482:2015wmo,BaBar:2020jma} and invisible decay channels~\cite{BaBar:2017tiz}, mainly by virtue of the $A'$ coupling to electrons and mesons. The searches utilized to probe these models, however, do not directly probe the interaction with muons which would be desirable to make robust claims about any BSM explanation
of \gmtwo~\cite{Chen:2017awl}. 

In this paper we consider searching directly for such a coupling using secondary particle beams at the proposed 
proton beam dump DarkQuest experiment~\cite{Apyan:2022tsd}. At this experiment, a high energy proton impinging on a target produces a large flux of secondary mesons. 
If these mesons decay into muons (like $\pi^\pm$ and $K^\pm$ do), they can produce muon-coupled force carriers in two ways: the 
new particle can be produced in the meson decays themselves, or via bremsstrahlung as the muons propagate through the beam dump.
Finally, some mesons can also decay into photons which can also produce these particles in photonuclear reactions via a muon-loop-induced coupling.

We illustrate the sensitivity 
of DarkQuest in the context of the muon-philic scalar model whose only tree-level interactions are with muons.\footnote{The sensitivity of DarkQuest to a variety of dark sector scenarios has been studied in Refs.~\cite{Gardner:2015wea,Berlin:2018pwi,Berlin:2018tvf,Dobrich:2019dxc, Tsai:2019buq, Batell:2020vqn, Blinov:2021say, Forbes:2022bvo,Ariga:2023fjg}. 
} Loop-suppressed couplings to photons are naturally expected 
and give rise to scalar decays when its mass is below the dimuon threshold. Thus the model's main signature at a beam dump like DarkQuest 
is a displaced decay into photons when the scalar is lighter than $2 m_\mu$. Ref.~\cite{Forbes:2022bvo} studied the complimentary case where the 
scalar is kinematically allowed to decay to muon pairs promptly. Our work compliments the existing literature on proposed fixed-target probes of 
muon-coupled particles at NA64, Fermilab, BDX, NA62, and SHiP~\cite{Gninenko:2014pea,Chen:2017awl,Berlin:2018pwi,Chen:2018vkr,Kahn:2018cqs,Marsicano:2018vin,Krnjaic:2019rsv,Rella:2022len,Cesarotti:2023udo}. 
Compared to previous studies of similar experimental setups and models (e.g.,~\cite{Berlin:2018pwi} focusing on a leptophilic scalar model at DarkQuest and~\cite{Rella:2022len} focusing on NA64 and SHiP), we improve the modeling of known production channels (muon bremsstrahlung) and identify additional relevant channels (meson decays, $K,\pi\to\mu\nu S$\footnote{These decays have also been studied in the context of kaon factories~\cite{Krnjaic:2019rsv}}), which turn out to dominate the sensitivity projections. We also provide the first more in-depth study of di-photon backgrounds at DarkQuest.

This work is organized as follows. In \cref{sec:darkquest} we briefly describe the DarkQuest experiment to establish key 
experimental parameters that determine its sensitivity (for more details see Ref.~\cite{Apyan:2022tsd}). We also estimate the fluxes of secondary particles that are relevant for signal and background production. 
We then specify the interactions of the muon-philic scalar model in \cref{sec:model_and_sim} and highlight the parameter space that can explain the \gmtwo anomaly. \cref{sec:signal_sim} details the modeling and simulation of signal events in three channels: muon bremsstrahlung, meson decays, and photon-induced processes. The compact nature of the DarkQuest experiment enables unique sensitivity to short BSM particle lifetimes at the cost of introducing several backgrounds in the case of photon signatures, which we discuss in \cref{sec:backgrounds}. There we identify processes that can mimic our signal and describe mitigation strategies. Unlike many other fixed-target searches for visibly-decaying particles, the sensitivity extends to large couplings without the characteristic exponential short lifetime cut-off.
Finally in \cref{sec:results} we present our sensitivity projections, showing that DarkQuest can decisively probe parts of the still-allowed \gmtwo parameter space.

We collect more details of our analysis in the appendices. Muon-philic scalar decays and certain production channels rely on the loop-induced scalar-photon coupling which is computed in \cref{sec:scalar_photon_coupling} (while the corresponding on-shell amplitude/form-factor is well known, we also provide a corrected off-shell amplitude). \cref{sec:vll_uv_completion} discusses a possible ultraviolet completion of the muon-philic scalar model based on vector-like leptons, which can be constrained with high energy colliders. \cref{sec:muon_transport} describes the propagation of muons through the DarkQuest dump which is a key aspect of our bremsstrahlung and muon-induced background calculations. In \cref{sec:meson_decay_details} we present the full expression for the three-body kaon decay partial width which dominates the signal reach.

%%%%%%%%%%%%%%%%%%%%%%%%%%%%%%%
\section{The DarkQuest Experiment}
\label{sec:darkquest}

%%%%%%%%%%%
\subsection{The Experimental Setup}
\begin{figure*}[tbh]
    \centering
    \includegraphics[width=0.75\textwidth]{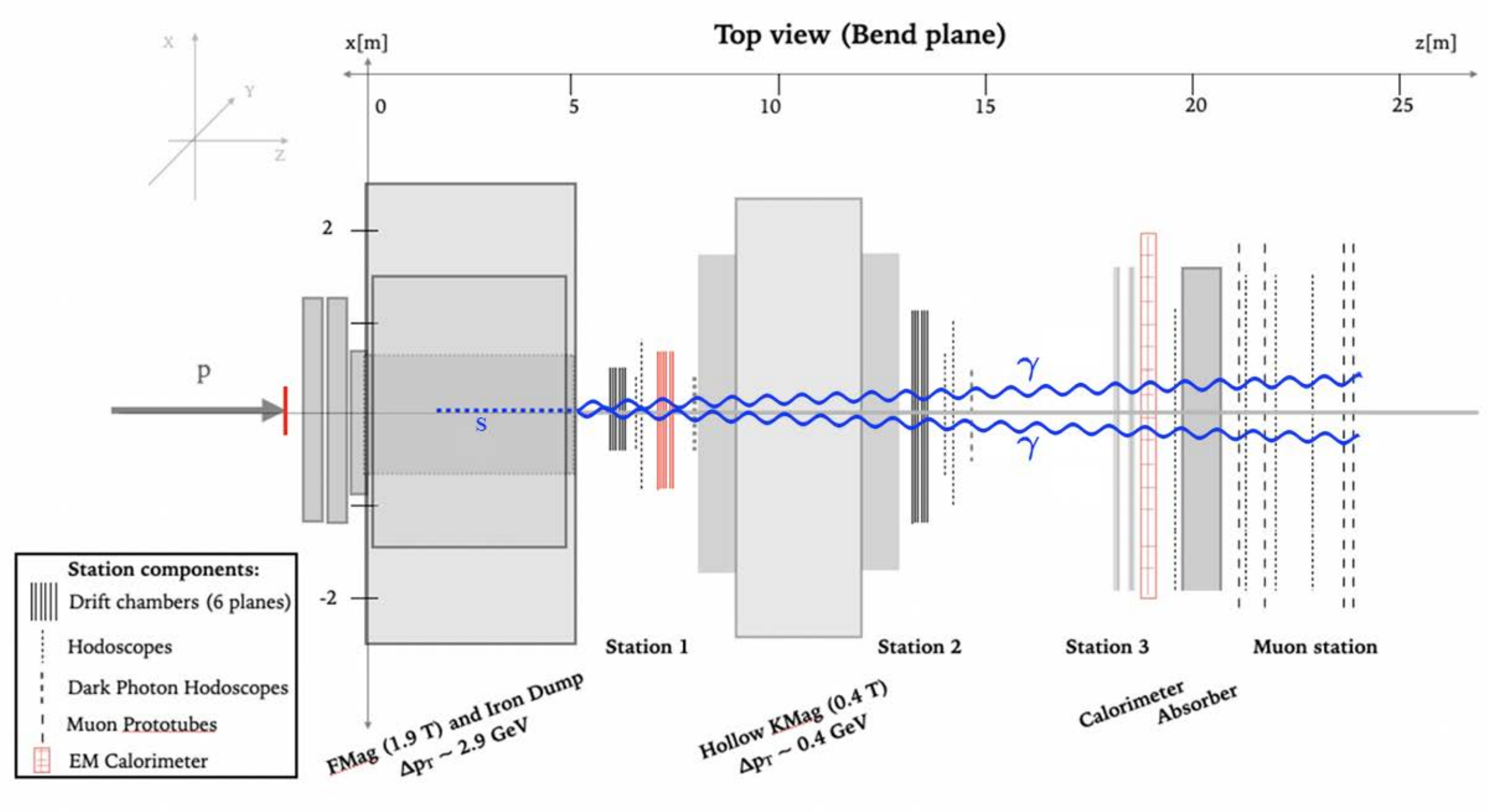}
    \caption{Side view of the DarkQuest experiment including proposed the ECal, target, and tracking detector upgrades, in red. A dark sector signature is illustrated in blue. Adapted from~\cite{Apyan:2022tsd}.}
    \label{fig:darkquest}
\end{figure*}
DarkQuest is a dark sector-focused upgrade of SpinQuest, a proton fixed-target spectrometer experiment on the neutrino-muon beamline of the Fermilab Accelerator Complex. This beamline supplies a high-intensity beam of 120 GeV protons from the Main Injector. Details of the set up are described in Refs.~\cite{SeaQuest:2017kjt} and~\cite{Apyan:2022tsd}. Here we summarize the key features relevant for our analysis. 

The setup of the DarkQuest experiment is shown in~\cref{fig:darkquest}. A 5 m long magnetized iron block (``FMAG'') is placed 1 m downstream from a thin nuclear target, serving as a focusing magnet and as a beam dump. Its magnetic field of $\sim 1.8$ T imparts a transverse kick of $\Delta p_T\sim 2.6$ GeV to ultra relativistic charge-one particles traversing the entire dump on average (the actual $p_T$ for individual particles is sensitive to multiple scattering and energy loss fluctuations in addition to the magnetic field, see~\cref{sec:muon_transport}). 
The FMAG stops most of the primary beam and secondary particles produced by proton-iron interactions, other than neutrinos and muons; despite this, the severely attenuated fluxes of protons and kaons can still be relevant for background processes. 

The detector extends for $\sim 20$ m after the dump and is composed of a series of tracking stations and an open-aperture magnet (``KMAG'') that further sweeps away charged SM radiation thanks to a transverse kick of $\Delta p_T\sim 0.4$ GeV. The above components are already present in the SpinQuest experiment. The DarkQuest upgrade will install: an electromagnetic calorimeter (``ECal'') at $\sim 19$ m between station 3 and the muon station, additional tracking layers to improve particle identification and hodoscopes for better triggering. We will also consider modifications of this set up that include more shielding before or after the FMAG. This is one strategy that can be used to suppress some (but not all) backgrounds as we discuss in \cref{sec:backgrounds}.

The DarkQuest experiment will collect $\sim 10^{18}$ protons on target (POT) in a two year long run. Depending on the status of the Fermilab accelerator complex \cite{Gori:2024zbs}, longer DarkQuest runs will be possible in the future. In our analysis, we will discuss the realistic scenario of $10^{18}$ POT and also a futuristic possibility in which DarkQuest will accumulate $10^{20}$ POT.

%%%%%%%%%%%
\subsection{Production of SM Particles}
The high intensity proton beam produces many SM particles at the beginning of the dump, including mesons, baryons, muons, and photons. These secondary particles open different production channels for the muon-philic scalar, and also produce backgrounds. In~\cref{tab:meson_counts_in_first_interaction_length}, we summarize the average counts of the relevant mesons and hadrons produced in the first proton interaction length $\lambda_{\rm{int}}= 16.77$ cm (approximately $1-e^{-1}\approx 0.63$ of the primary proton beam interacts within this region). We estimated these yields using \pythia 8.306~\cite{Bierlich:2022pfr} \footnote{In the table and in the discussion afterwards, we do not report the production rates for charmed mesons. Charmed mesons contribute to the production of muon-philic scalars but they are sub leading to kaons.}. 

Short-lived mesons like $\pi^0$ decay promptly after production. $\pi^0$ along with $\eta^{(\prime)}$ decays generate a flux of photons at the beginning of the FMAG. As we will discuss, decays of the long-lived mesons $\pi^\pm$ and $K^\pm$ contribute to signal production in two ways: first, through rare three body decays, and second, by generating a flux of muons that can undergo bremsstrahlung.\footnote{Muons are also produced in the Drell-Yan process, but the corresponding cross-section is smaller than the total $pp$ cross-section by a factor of $\sim 10^{7}$, so this mechanism is subdominant to production from meson decays.}  We focus on production and decay of these mesons in the first meson interaction length for simplicity. 
This allows us to neglect energy losses of hadrons that decay or interact deep in the dump. We expect this to be a good $\mathcal O(1)$ approximation for the signal. This is because while including a longer decay region for $\pi^\pm$, $K^\pm$ would increase the muon and scalar yields linearly with the decay region size, the meson intensity falls exponentially with depth. Therefore decays in the first interaction length account for the majority of the (scalar or muon) flux that is relevant for the signal and background estimates. 
We estimate the muon and photon yield, and the number of $\pi^\pm$ and $K^\pm$ decays in the first interaction length in~\cref{tab:muons_photons_and_decays_in_first_interaction_length}. 
The latter counts are obtained via 
\beq
N_{M,\;\mathrm{dec}} \simeq \Npot n_M \Gamma_M \langle \gamma_M^{-1}\rangle \lambda_M\,,
\label{eq:number_of_prompt_decays}
\eeq
where $\Npot$ is the number of POT, $n_M$ is the meson yield per POT in the first interaction length (from \cref{tab:meson_counts_in_first_interaction_length}), $\Gamma_M$ and $\lambda_M$ are the meson decay rate and interaction length, respectively. The average over inverse boosts $\langle \gamma_M^{-1}\rangle$ is performed using Monte Carlo samples from \pythia. 

For certain backgrounds the attenuated proton beam or secondary $K_L$ and $\Lambda$ fluxes \emph{are} an important source, and we will comment on them in~\cref{sec:backgrounds}. 

  \begin{table}
    \begin{tabular}{|L|L|L|L|L|L|L|L|}
      \hline
      \pi^0 & \eta & \eta^\prime & \pi^\pm  & K^\pm & K_L & K_S & \Lambda\\
      \hline
      1.5 & 0.19 & 0.021 & 3.2 & 0.25 & 0.11 & 0.21 & 0.082\\ 
    \hline
  \end{tabular}
  \caption{Average number of mesons or baryons produced in the first interaction length of the FMAG per 120 GeV proton on target. These counts are estimated using \texttt{Pythia}.
  \label{tab:meson_counts_in_first_interaction_length}}
  \end{table}
  \begin{table}
    \begin{tabular}{|L|L|L|L|}
      \hline
      \mu^\pm & \gamma & \text{prompt } \pi^\pm \text{ dec.}  & \text{prompt } K^\pm \text{ dec.}\\
      \hline
      1.0\times 10^{16} & 3.8 \times 10^{18} &  8.6\times 10^{15} & 1.9\times 10^{15}\\
    \hline
  \end{tabular}
  \caption{Average counts for particles that are produced or decay near the front of FMAG for $10^{18}$ POT. The first two columns contain the number of muons and photons produced in the first proton interaction length of the FMAG, while the last two columns contain the number of prompt $\pi^\pm$ and $K^\pm$ decays as estimated via \cref{eq:number_of_prompt_decays}. Here ``prompt'' refers to decays within the first meson interaction length. 
    \label{tab:muons_photons_and_decays_in_first_interaction_length}
  }
  \end{table}

  %%%%%%%%%%%
\subsection{Di-photon mass resolution}
\label{sec:diphoton_mass_resolution}
As we will discuss in \cref{sec:backgrounds}, di-photon signatures are not background-free at DarkQuest. The most important backgrounds is $\pi^0$ and $\eta$ production at the end of the dump produced either from the attenuated proton beam or from the secondary muon beam undergoing deep inelastic scattering. One experimental handle that can disentangle these 
events from the signal are selections on diphoton invariant mass. Having a good mass resolution will improve the discrimination between signal and background. The resolution can be estimated using the energy and position resolution of the ECAL~\cite{Apyan:2022tsd}. In particular, we take a position resolution of 3 cm and an energy resolution of $1\%+7\%/\sqrt{E(\rm{GeV})}$. 

By simulating $\pi^0$, $\eta^{(\prime)}$ production and their decays at the back of the dump we find that the mass resolution varies between $10\%$ and $30\%$ for typical meson energies, with a better resolution for the $\eta^{(\prime)}$, if compared to the $\pi^0$.\footnote{We thank Yongbin Feng for useful discussions regarding these calculations.} We will find that even taking the optimistic $10\%$ figure, selections of diphoton invariant mass can reduce background rates by a factor of a few, but not completely eliminate them. Other mitigation strategies will still be required (see \cref{sec:results}).

%%%%%%%%%%%%%%%%%%%%%%
\section{The Scalar Singlet Model}
\label{sec:model_and_sim}

%%%%%%%%%%%%%%%%%%%
\subsection{The Lagrangian of the Muon-philic Scalar}
\label{sec:model}

We focus on minimal scalar model that at the renormalizable level contains only the interaction with 
muons 
\beq
\mathscr{L} \supset \frac{1}{2}(\partial S)^2 - \frac{1}{2} m_S^2 S^2 
- g_S S \bar{\mu } \mu.
\label{eq:lang}
\eeq
We emphasize that this scenario does not have a coupling to electrons at tree level; in the parlance of Ref.~\cite{Chen:2017awl} 
this is ``Model B''.
The effective theory also contains interactions with photons at dimension five
\beq
\mathscr{L} \supset - \frac{1}{4} g_{S\gamma\gamma} S F_{\mu\nu} F^{\mu\nu}.
\label{eq:photon_coupling}
\eeq
The coupling $g_{S\gamma\gamma}$ receives contributions from ultraviolet (UV) physics and from loops of muons:
\beq
g_{S\gamma\gamma} = g_{S\gamma\gamma}^{(\mathrm{UV})} + \frac{\alpha g_S}{2\pi m_\mu} f_{1/2}\left(\frac{4m_\mu^2}{m_S^2}, 0\right),
\label{eq:photon_coupling_uv_and_ir}
\eeq
where in writing the loop function $f_{1/2}$ we took $S$ and the photons to be on-shell. 
We will neglect the UV contribution in what follows in order to be consistent with previous studies. The usual explanation given for this is that one 
naively expects $g_{S\gamma\gamma}^{(\mathrm{UV})}$ to be smaller by a factor of $m_\mu/M$ where $M$ is the UV scale, see \cref{sec:scalar_photon_coupling}.
However, in realistic UV completions, the $S$ coupling to muons (i.e., $g_S$) and the heavy new physics scale, $M$, are independent, so the UV and infrared (IR) contributions to $S\gamma\gamma$ can potentially compete for some choices of parameters, as discussed in \cref{sec:vll_uv_completion}.
This photon interaction gives rise to $S$ decays into photons with rate:
\beq
\Gamma_S = \frac{\alpha^2 g_S^2 m_S^3}{256\pi^3 m_\mu^2} \left|f_{1/2}\left(\frac{4m_\mu^2}{m_S^2},0\right)\right|^2,
\eeq
where the loop function $f_{1/2} \to 4/3$ as $m_S/m_\mu \ll 1$ (the full result is given in \cref{sec:scalar_photon_coupling}). The muon loop also generates couplings 
of off-shell photons to $S$ which are relevant for, e.g., the photon fusion and Primakoff $S$ production. The $S\gamma\gamma^*$ amplitude needed for this calculation is given in \cref{sec:scalar_photon_coupling}. 

Clearly the model in~\cref{eq:lang} is not electroweak gauge-invariant. From the point of view of UV completions, the inclusion of couplings 
to electrons and taus is natural in minimal models with Higgs portal interactions or extra Higgs doublets above the electroweak scale~\cite{Batell:2016ove}. 
However, these models have recently been excluded as solutions of \gmtwo~\cite{BaBar:2020jma}.
The \emph{flavor-specific} interaction in~\cref{eq:lang}, while more minimal in the infrared, requires more engineering in the UV. 
Ref.~\cite{Batell:2017kty} delineated the conditions under which couplings to specific SM fermion mass eigenstates (the muon in our case) are obtained in a 
technically-natural and experimentally-viable way. 
As a first step, the flavor-specific coupling to muons can be obtained from the electroweak-invariant effective operator 
\beq
- \frac{S}{M} H^\dagger L \mathbf{c}_S \ell_R^c + \hc,
\label{eq:uv_completion1}
\eeq
where $H$ ($L$) is the Higgs (lepton) $SU(2)_L$ doublet, $\ell_R^c$ is the $SU(2)_L$ singlet lepton (in Weyl notation), 
$M$ is a UV scale, and $\mathbf{c}_S$ is a matrix in flavor space.
In order for this interaction to reduce to~\cref{eq:lang}, 
$\mathbf{c}_S$ has to obey two conditions: 1) it must be diagonal in the lepton mass basis (i.e., $\mathbf{c}_S$ is simultaneously diagonalizable with 
the lepton Yukawa matrix) and 2) $\mathbf{c}_S = \diag(0,c_S,0)$ in the charged lepton mass basis. This structure generates 
the desired coupling to muons with $g_S = c_S v/(\sqrt{2}M)$, where $v=246$ GeV, and it minimizes flavor-changing neutral currents. In general, 
$c_S$ is complex, but its imaginary part shifts \gmtwo in the wrong direction~\cite{Batell:2017kty}. We therefore focus on purely real $c_S$. 
The spurion analysis of Ref.~\cite{Batell:2017kty} reveals that the intricate structure of $\mathbf{c}_S$ is radiatively stable.\footnote{The parity $S\to -S$ ensures that corrections to $c_S$ are always proportional to $c_S$, while any off-diagonal elements can only arise from the non-diagonal neutrino mass matrix.}  

The electroweak-invariant interaction in \cref{eq:uv_completion1} is still non-renormalizable and therefore requires a UV completion. This is easily achieved by adding a pair of vector-like leptons with mass $M$~\cite{Batell:2017kty}, in which case $\mathbf{c}_S$ is proportional to the Yukawa couplings of these new particles with the SM Higgs and with $S$. We explore such a UV completion in \cref{sec:vll_uv_completion}, which allows us to connect the low energy muon and photon couplings to UV parameters, and to consider naturalness and direct experimental bounds that must be satisfied.
This scenario still does not explain the flavor alignment of $\mathbf{c}_S$ and the SM lepton Yukawa matrices.

%%%%%%%%%%%%%%%%%%%

\subsection{The New Physics Contribution to \boldmath{$(g-2)_\mu$}}
\label{sec:gm2_from_muon-philic_scalar}
The model in \cref{eq:lang} is one of the few available minimal explanations of the $(g-2)_\mu$ anomaly with new physics at or below the several GeV scale \cite{Capdevilla:2021kcf}. The contribution of $S$ to \gmtwo is ~\cite{Batell:2016ove}
\begin{equation}
  \Delta a_\mu^{S} = \frac{g_S^2}{8\pi^2} \int_0^1 dx \frac{(1+x)(1-x)^2}{(1-x)^2 + x (m_S/m_\mu)^2}.
\end{equation}
The anomaly in \cref{eq:g_minus_two_discrepancy} is explained if $g_S\sim 4\times 10^{-4}$ for $m_S \lesssim 2m_\mu$. 
\bigskip

%%%%%%%%%%%%%%%%%%%

\section{Signal Production and Simulations}
\label{sec:signal_sim}
\begin{figure*}[t]
  \centering
  \includegraphics[width=0.9\textwidth]{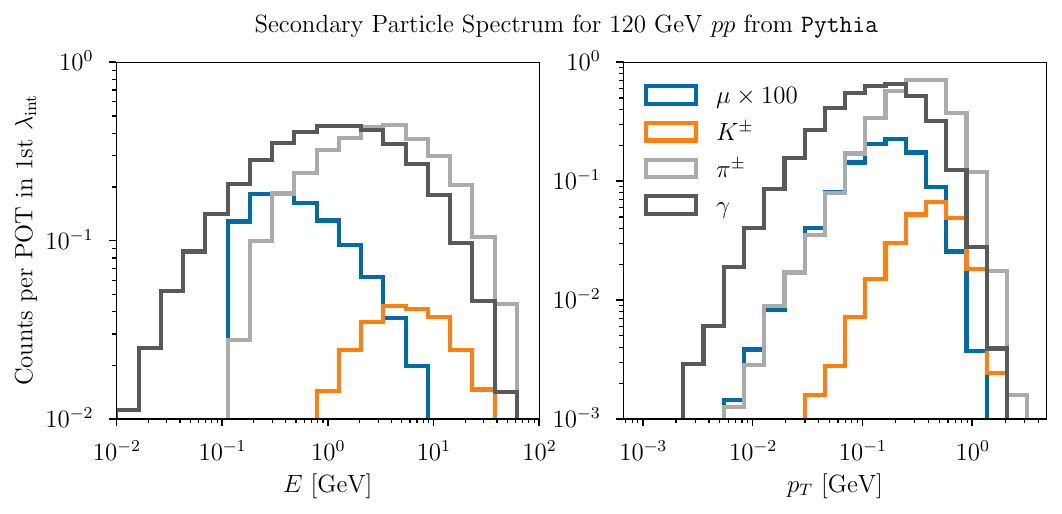}
  \caption{Counts of secondary particles per 120 GeV proton on target as a function of energy (left panel) and transverse momentum (right panel). The $\pi^\pm$, $K^\pm$ ($\gamma$, $\mu^\pm$) spectra sum to the average yields in the first interaction length of the dump given in \cref{tab:meson_counts_in_first_interaction_length} (\cref{tab:muons_photons_and_decays_in_first_interaction_length}, divided by $10^{18}$). Note that the muon spectrum includes the probability for the parent particles (mostly $\pi^\pm$ and $K^\pm$) to decay in the first interaction length (this spectrum is also multiplied by 100 for convenient visualization).\label{fig:secondary_distributions_from_120_GeV_pp}}
\end{figure*}

The dark scalar can be produced in several processes enabled by secondary muons, mesons and photons present in the beam dump. 
The energy and transverse momentum spectra of these particles are shown in \cref{fig:secondary_distributions_from_120_GeV_pp}.
We find that muon bremsstrahlung and meson decays dominate the dark scalar yield in interesting 
regions of parameter space, while the rate of photon-initiated reactions is loop-suppressed. We discuss these processes in more detail below. 

\subsection{Muon Bremsstrahlung}
\begin{figure}
  \centering
    \includegraphics[width=0.48\textwidth]{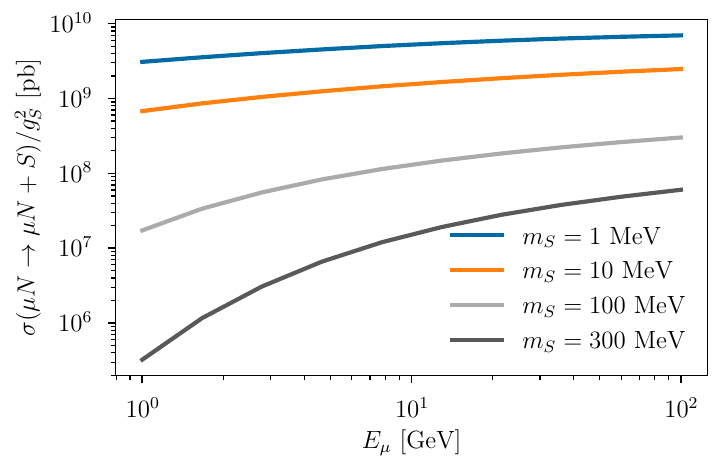}
    \caption{Scalar bremsstrahlung ($\mu N\to \mu N S$) cross-section, in iron as a function of muon energy for several choices of $m_S$. Note that to obtain a realistic scalar yield at DarkQuest for this process one must convolve these cross-sections against the muon spectrum and account for muon propagation effects. \label{fig:scalar_brem_xsec}}
\end{figure}

The primary proton collisions with the dump lead to a large flux of secondary muons. These muons are generated from meson decays ($\pi^\pm$ and $K^\pm$ in particular), and from Drell-Yan production. These muons can then interact within the dump, radiating a scalar. The scalar bremsstrahlung cross-section has been computed 
exactly in~\cite{Tsai:1986tx,Tsai:1989vw,Liu:2016mqv}, but most recent works use the Weiszacker-Williams (WW) approximation, see, e.g., Ref.~\cite{Chen:2018vkr,Rella:2022len}. This cross-section enters the signal prediction calculation in two important ways: first, it normalizes the overall rate and, second, its 
differential is used to generate $S$ kinematics. We have compared three ways of evaluating the overall cross-section: (1) using the WW approximation; (2) numerically computing it in \texttt{MadGraph}~\cite{Alwall:2014hca}; (3) integrating the exact results of Ref.~\cite{Liu:2016mqv} using \texttt{VEGAS}~\cite{Lepage:1977sw,Lepage:2020tgj}. 
In each case we implemented the coherent bremsstrahlung cross-section following the references above and used the elastic and inelastic nuclear and atomic form-factors from Ref.~\cite{Kim:1973he,Bjorken:2009mm}.\footnote{Note that \cite{Bjorken:2009mm} has a small typo in the nuclear inelastic form factor which was pointed out in Refs.~\cite{Jodlowski:2019ycu,Celentano:2020vtu}.}
In principle the latter two methods should give very similar results, but in practice we find that the custom \texttt{VEGAS} implementation yields results that are more numerically stable.\footnote{The difficulty of simulating light-particle bremsstrahlung using off-the-shelf Monte Carlo tools has been noted in, e.g., ~\cite{Celentano:2020vtu,Forbes:2022bvo}.} The result of this calculation is illustrated in \cref{fig:scalar_brem_xsec} for several values of $m_S$ - we find excellent agreement with results of Ref.~\cite{Marsicano:2018vin} which used a different numerical method, which is also based on the integration of the full matrix element squared. For the main results in this work we use the \texttt{VEGAS} calculation of the total cross-section, noting that the WW approximation predicts scalar yields that are a factor of 2-5 larger for $m_S \lesssim 2m_\mu$ (such discrepancies were also observed in Refs.~\cite{Liu:2016mqv,Rella:2022len}). We will use \texttt{MadGraph} to generate final state scalar and muon kinematics. 

Most of the muons are produced near the front of the dump, so they can propagate a significant distance before undergoing bremsstrahlung. During this propagation 
they experience energy losses, multiple Coulomb scattering, and their trajectory is bent in the magnetic field. There have been two different approaches to modeling these effects. First, input muon spectrum can be used to predict the signal yield by convolving it with the differential cross-section~\cite{Berlin:2018pwi,Rella:2022len} 
\begin{widetext}
\begin{align}
\label{eq:NsignalMuon}
N_\text{signal} \simeq n_\text{atom}  \int d E_\mu^0 \frac{d N_\mu}{d E_\mu^0}  \int_{E_\mu^\text{min}}^{E_\mu^0} d E_\mu  \frac{\sigma_\text{Brem.} (E_\mu)}{|d E_\mu / d z_\mu|}  \int_{z_\text{min} - z_\mu}^{z_\text{max} - z_\mu}  d z \frac{1}{N_\text{MC}}\sum\limits_{\text{events $\in$ geom.}} \frac{e^{- z/(\gamma_S c\tau_S)}}{ \gamma_S c\tau_S}.
\end{align}
\end{widetext}
where $z_{\rm{min}}\simeq 5$ m, $z_{\rm{max}}\simeq 19$ m and $N_{\rm{MC}}$ is the total
number of simulated events. The sum is performed over only those events that remain within the geometry of the DarkQuest spectrometer. Here $E_\mu^0$
is the initial energy of the muon at production, and $E_\mu$ is the muon energy after traversing some finite distance, $z_\mu$, in the FMAG before radiating the scalar; $\gamma_S$ and $\tau_S$ are the boost and the lifetime of the scalar, respectively; $d N_\mu/d E_\mu^0$ is the total number of muons produced per initial energy bin within the first meson interaction length (output of \pythia); $n_{\rm{atom}}=8.5\times 10^{22}$/cm$^3$ is the number density of target iron nuclei and $\sigma_\text{Brem.}(E_\mu)$ is the muon bremsstrahlung cross-section, as a function of the muon energy (see Fig. \ref{fig:scalar_brem_xsec}). Note that in this approach muon propagation is simply encoded in the $dE_\mu$ integration, and the effect of a magnetic field is applied ``by hand'' as a shift in $p_T$ of each muon. Multiple scattering is not included.

Alternatively, each muon can be tracked through the dump using its equations of motion, until it undergoes a bremsstrahlung event. At this point $S$ kinematics is sampled from the differential cross-section. This approach makes it easier to include the effects of multiple Coulomb scattering, energy loss fluctuations, and magnetic fields. In Ref.~\cite{Marsicano:2018vin} this method was implemented within \texttt{GEANT4}~\cite{GEANT4:2002zbu,Allison:2006ve,Allison:2016lfl}. Here we follow a similar approach, but instead model the propagation in a custom \texttt{Python} program described in \cref{sec:muon_transport}. 

Finally, we note that muon bremsstrahlung in a thick target has the peculiar feature that scalar particles can be produced anywhere in the dump with similar probability, including very close to the end ($z_\mu\sim z_{\rm{min}}$). This can be easily seen from \cref{eq:NsignalMuon} in the limit of large muon energy (such that one can neglect energy losses) and short scalar lifetime ($\gamma_S c\tau_S \ll \zmin$). In this case the probability of the scalar to be produced in the dump and decay beyond it is approximately  
\begin{equation}
  \begin{split}
  n_{\mathrm{atom}} \sigma_\text{Brem.}(E_\mu) \int_0^{\zmin} d z_\mu e^{-(\zmin-z_\mu)/(\gamma_Sc\tau_S)} \\
  \approx n_{\mathrm{atom}} \sigma_\text{Brem.}(E_\mu) \gamma_S c \tau_S, 
\end{split}
\end{equation}
where we assumed that the cross-section and energy loss rates are constant. This shows that the signal yield will be dominated by the scalar production in the last $S$ decay length of the shield. In contrast, if the scalar was produced near the front of the dump, the rate would be proportional to $e^{-\zmin/(\gamma_S c\tau_S)}$. 
Therefore the projected sensitivity of this channel extends to large couplings, which is qualitatively different to many other long-lived-particle (LLP) searches which feature a very thick shield. For example, the proton beam-dump experiment CHARM~\cite{CHARM:1983ayi,CHARM:1985nku,Gninenko:2012eq} had a shield thickness of $\sim 480$ m so that the LLP signal in the large coupling/short lifetime regime is exponentially suppressed despite the higher beam energy of $400$ GeV. Note that potential signals from secondary muons (which can propagate over such distances) have not been studied, but the muons would likely experience significant energy loss, making it difficult for signals to pass detector selections.

\subsection{Meson Decays}
Dark scalars are also produced from the charged current meson decays $\pi^\pm\to\mu^\pm\nu S$ and $K^\pm\to\mu^\pm\nu S$. 
Using the leading-order chiral perturbation theory Lagrangian
\begin{equation}
  \mathscr{L}_{\chi\mathrm{PT}}\supset \sqrt{2} G_F f_P V_{q_1 q_2} (\partial_\alpha  P^-) (\bar \mu \gamma^\alpha P_L \nu_\mu) + \hc,
\end{equation}
we find the partial width for meson $P = \pi,K$
\begin{subequations}
\begin{align}
  \Gamma(P^\pm\to \mu^\pm \nu S) & \approx \frac{g_S^2 f_P^2 G_F^2 |V_{q_1 q_2}|^2 m_P^3}{768\pi^3}\\
                         & \approx \frac{g_S^2 m_P^2}{96\pi^2 m_\mu^2} \Gamma(P^\pm\to \mu^\pm\nu),
\end{align}
\end{subequations}
in the $m_P\gg m_S, m_\mu$ limit (the full expression is given in~\cref{sec:meson_decay_details}). 
Here $f_P$ is the meson decay constant ($f_\pi = 130\;\MeV$ and $f_K= 156\;\MeV$ in our conventions); $V_{q_1 q_2}$ is the relevant CKM matrix element: $|V_{ud}|\approx 0.9737$ ($|V_{us}|\approx 0.2246$) for $\pi$ ($K$) decay. These are the most important meson production modes for the scalar thanks to their sizable branching ratio into a muon and a neutrino.

An additional production mode is $K_L\to \pi^\pm \mu^\mp\nu_\mu S$. The contribution from this process, however, is generically smaller than from the meson three body decays discussed above.

\subsection{Photon-induced Processes}
Photons from meson decays and electromagnetic showers can produce $S$ through several reactions. These include 
  the tree-level process $\gamma + N \to S + \mu^+ + \mu^- + N$,\footnote{We thank the anonymous referee for bringing this process to our attention.}
  Primakoff process $\gamma + N \to S + N$ and photon fusion $\gamma^* \gamma^* \to S$, 
where $\gamma^*$ are virtual photons corresponding to the electromagnetic fields of a beam proton and target nucleus. 
The former reaction is mediated by the tree-level $S$-muon coupling, while the latter two are due to the scalar-photon coupling in \cref{eq:photon_coupling} 
and are therefore analogous to axion-like particle production in beam dumps. 

For the tree-level process we computed the photon spectrum-averaged cross-section using \texttt{MadGraph}. We found that throughout the mass range of interest 
the resulting $S$ yield is at least 2 orders of magnitude smaller than from muon bremsstrahlung and meson decay. 
%The scalar-photon interaction in \cref{eq:photon_coupling} enables several production mechanisms that are analogous to axion-like particle 
%production in beam dumps. These include the Primakoff process $\gamma + N \to S + N$ and photon fusion $\gamma^* \gamma^* \to S$, 
%where $\gamma^*$ are virtual photons corresponding to the electromagnetic fields of a beam proton and target nucleus. 
Primakoff production 
was shown to dominate over photon fusion in thick targets due to the large secondary photon flux~\cite{Blinov:2021say}. 
The scalar-photon interaction needed to evaluate the Primakoff cross-section is described in~\cref{sec:scalar_photon_coupling}. A potential difference of the present model 
with respect to axion-like particles is that the photon coupling is generated by muons, leading to a non-trivial energy dependence of that coupling (because the muon mass is not 
much larger than other experimental energy scales). We have estimated the yield of dark scalars from this process, finding that it is 1--2 orders of magnitude smaller than for bremsstrahlung in the mass range of interest.\footnote{Moreover, the corresponding signal would not have a on-shell muon in the final state which may be an important event selection designed to reduce diphoton backgrounds as we discuss in the next two sections.}
The relative importance of the tree-level process and Primakoff is a function of mass, with Primakoff dominating above $m_S \approx 30\;\MeV$. However, both reactions are negligible sources of $S$ compared to bremsstrahlung and meson decay throughout our parameter space.

\section{Backgrounds}
\label{sec:backgrounds}
The compact geometry of DarkQuest enables searches for particles with short decay lengths (especially if they are radiated by muons propagating through the dump). This feature leads to a complementary reach compared to other experiments with a larger dump and smaller angular acceptance. The comparatively short shield also means that potential signals must contend with backgrounds.
There is a variety of SM processes that can mimic the production and decay of a BSM 
long-lived particle into photons. 
The three types of potentially problematic SM events are: 1) production of genuine SM long-lived particles, such as $K_L$; 2) neutral mesons ($\pi^0$, $\eta^{(\prime)}$) produced at the very back of the dump from the attenuated proton beam; and 3) production of neutral mesons ($\pi^0$, $\eta^{(\prime)}$) via Deep-Inelastic Scattering (DIS) of muons towards the end of the dump. We will show that the events of the third kind are the most dangerous, but 
that they can be mitigated with several event selections, while preserving experimental coverage of the most interesting \gmtwo band at $m_S \lesssim 2m_\mu$. These backgrounds and potential mitigation strategies are summarized in \cref{tab:background_summary} -- they will be discussed in more detail below and in \cref{sec:results}.

  \begin{table*}
    \begin{tabular}{|c|c|| c|c|c|c|c|}
      \hline
      & Raw counts & ECAL hit & $\gamma$ sep. & $\mu$ hit & $m_{\gamma\gamma}$ & Extra Shielding \\
      \hline
      $p$-induced $K_L$ & $6\times 10^4$ & $\checkmark$ & $\checkmark$ & $\checkmark$ & $\checkmark$ & $\checkmark$ \\
      $p$-induced $\pi^0,\; \eta^{(\prime)}$ & $2\times 10^4$ & $\checkmark$ &  & $\checkmark$& $\checkmark$ & $\checkmark$\\
      $\mu$-DIS-induced $\pi^0,\; \eta^{(\prime)}$ & $10^6$ & $\checkmark$ & & & $\checkmark$ & \\
      \hline
  \end{tabular}
  \caption{Summary of SM background events that can mimic the signal $S\to \gamma \gamma$ and experimental techniques that can reduce or eliminate them. The first two rows correspond to mesons produced by the primary proton beam. 
  The last row corresponds to mesons produced deep in the dump by a secondary muon beam via deep-inelastic-scattering. The columns to the right of the double-line divider indicate whether the listed background events can be significantly reduced using the corresponding experimental handle. ``ECAL hit'' requires at least one photon with a sufficient energy in the ECAL; ``$\gamma$ sep.'' corresponds to exactly two well-separated photons in the ECAL; ``$\mu$ hit'' is the requirement of detecting a muon; ``$m_{\gamma\gamma}$'' is a selection based on the diphoton invariant mass potentially measurable with the ECAL; ``Extra shielding'' corresponds to adding additional material behind the FMAG. The impact of these selections on the background yield is quantified in \cref{sec:backgrounds} and \cref{sec:results}.}
  \label{tab:background_summary}
  \end{table*}

\subsection{SM Long-Lived Particles}
\label{sec:sm_llps}
Standard Model LLPs, such as $K_L$ and $\Lambda$, which decay into neutral pions can 
produce signal-like photons. We will focus on $K_L$ decays in the fiducial region $z > 5$ m and $z <19$ m (before the ECal). $\Lambda$'s have a shorter attenuation 
length compared to mesons, so their contribution to background ends up being two orders of magnitude smaller. The number of LLP background 
events can be estimated as
\begin{equation}
  \begin{split}
    & n_{K_L} \Npot \mathrm{BR}(K_L\to 3\pi^0) e^{-z_{\mathrm{shield}}/\lambda_{K\;\mathrm{int}}} \\
    & \times  \left[e^{-z_\mathrm{shield}/(\gamma c\tau_{K_L})}  - e^{-z_\mathrm{max}/(\gamma c\tau_{K_L})} \right]  \\
    &   \approx 6\times 10^4,
\end{split}
\label{eq:KL_yield_estimate}
\end{equation}
where $n_{K_L}$ is the number of $K_L$ produced per POT from \cref{tab:meson_counts_in_first_interaction_length}, $\mathrm{BR}(K_L\to 3\pi^0)\approx 0.2$ is the branching fraction of the most important $K_L$ decay channel that gives photons in the final state, $z_{\mathrm{shield}}=5$ m is the length of shielding, $\lambda_{K\;\mathrm{int}} \approx 20\;\mathrm{cm}$ is the kaon interaction length in iron. The factor on the second line is the probability of the kaons to decay after the shielding and before the ECAL. The number of events in the last line is computed for $N_p=10^{18}$ POT. Each of these events can lead to a $\leq 6$ photons signal in our detector. Note that most $K_L$'s are produced near the front of the dump as the incoming proton beam is exponentially attenuated.\footnote{$K_L$'s produced deeper in the dump experience less attenuation before they exit which somewhat off-sets the attenuation of the beam that produced them, compared to $K_L$'s near the front. However, because the kaon interaction length is longer than the proton interaction length, the latter $K_L$ flux still dominates.}

There are two approaches to deal with this background. First, one can use additional shielding in front or behind the FMAG (``Extra shielding''  in~\cref{tab:background_summary}). 
Using \pythia to estimate the median meson production rates and boosts, we find that an additional 2 m of (iron) shielding would bring the $K_L$ 
background rate below 10 events for $10^{18}$ POT. Similar shielding requirements were discussed in Refs.~\cite{Berlin:2018pwi,Blinov:2021say} in the context of axion-like particle searches. 
It is important to note that if the only goal is to close the \gmtwo window, we can significantly relax the shielding requirement.
This is because we expect $\mathcal{O}(10^3- 10^6)$ signal events in the currently-allowed \gmtwo band at $m_S\lesssim 2m_\mu$ (see \cref{fig:event_yield_contours_g_minus_two_region}).
The amount of extra shielding needed to bring the background below $10^3$ events is only $90$ cm. This can be further reduced by using a different material, such as tungsten. 

The second approach to reducing SM LLP backgrounds is to select events with exactly two photons in the ECAL and a single muon hit (``$\gamma$ sep.'' and ``$\mu$ hit'' in~\cref{tab:background_summary}). Both the kaon decay and muon bremsstrahlung $S$ production mechanisms can in principle pass these cuts, while the backgrounds either have too many photons, do not have a muon in the final state or both. The leading $K_L$ decay that leads to background events is $K_L \to 3\pi^0 \to 6\gamma$, which would be completely eliminated by such cuts, without requiring any additional shielding. In~\cref{sec:results}, we will show that enough signal events pass these selections to retain sensitivity to the open \gmtwo region at $m_S\lesssim 2m_\mu$.

\subsection{Secondary Mesons from the Back of the Dump}
\label{sec:mesons_from_back_of_the_dump}
Another type of potentially problematic SM events is the production of $\pi^0$ and $\eta^{(\prime)}$ from the
attenuated proton beam in the last radiation length of the dump. The prompt photons from their decays 
would have a large probability of escaping the dump and potentially ending up in the ECAL. The number of such events is approximately 
\begin{equation}
\begin{split}\label{eq:LLPBack}
&  (n_{\pi^0}+n_{\eta} + n_{\eta^\prime}) \Npot e^{-z_{\rm{shield}}/\lambda_{p\; \rm{int}}}\frac{\lambda_{\rm{rl}}}{\lambda_{p\;\rm{int}}}\\
  &~~~~\approx 2\times 10^{4},
\end{split}
\end{equation}
where $\lambda_{p\;\rm{int}}$ = 16.77 cm is the nuclear interaction length for protons,
$n_{M}$ are given in \cref{tab:meson_counts_in_first_interaction_length} and $\lambda_{\rm{rl}}$= 1.757 cm is the radiation length in iron.\footnote{The attenuated proton beam looses $\sim 10$ GeV of energy as it propagates through the FMAG, but the meson yields at this lower energy are approximately the same as for in original 120 GeV beam.} 
The last factor, $\lambda_{\rm{rl}}/\lambda_{p\;\rm{int}}$, 
is the probability for protons to interact in the final radiation length of the dump. 
The background events in the last line of~\cref{eq:LLPBack} can be eliminated in a few different ways. 
Thanks to the exponential suppression in~\cref{eq:LLPBack}, an additional 1.8 m (70 cm) of iron shielding would suffice to reduce this background to below 1 ($10^3$) event for $\Npot = 10^{18}$. 
Alternatively, as for SM LLPs, event selections requiring two photon hits in the ECAL and a muon hit would also help thanks to a requirement on the di-photon invariant mass and distributions, see~\cref{sec:results}. We will investigate how such selections 
affect the signal yield in the next section. 

\subsection{Mesons from $\mu$ DIS}
\label{sec:mu_dis}
Finally, we consider backgrounds generated by muons undergoing DIS at the end of the dump and producing $\pi^0$ and other mesons that can decay to photons. The reason why this reaction chain is potentially important is because muons are not exponentially attenuated by the dump, unlike the original proton beam (c.f., the secondary meson background in~\cref{sec:mesons_from_back_of_the_dump}). In order to estimate the rate, we first produce a distribution of muons at the end of the dump as follows.   
We simulate the primary $pp$ collisions in \pythia that produce $\pi^{\pm}$ which are then 
decayed to muons in the first interaction length. The muons are then propagated to the back of the dump as described in \cref{sec:muon_transport}. We find that 
the number of muons at the back of the 5 m dump with sufficient energy to initiate DIS is $n_\mu^{(\mathrm{back})}\sim 5\times 10^{-6}$ per $pp$ interaction. Note that this number is not sensitive to the dump thickness because the muons are not strongly attenuated. This is because the muons lose only about $\sim 10\;\MeV/\mathrm{cm}$, 
so even an additional 2 m of iron does not drastically affect their spectrum at the back of the dump.

The propagated sample of muons is then fed into \pythia to simulate DIS, $\mu p \to \mu + X$, which is used to estimate the 
meson-count weighted cross-sections, $\langle \sigma(\mu p \to \mu + X) n_M\rangle$ where $M = K_L,\; \pi^0$. 
We find that 
\begin{equation}
  \begin{split}\label{eq:DISCross}
    \langle \sigma(\mu p \to \mu + X) n_{\pi^0}\rangle & = 2\times 10^{-5}\;\mathrm{mb}, \\
    \langle \sigma(\mu p \to \mu + X) n_{K_L}\rangle & = 5\times 10^{-7}\;\mathrm{mb}.  
\end{split}
\end{equation}

We can now put all of these ingredients together. First, the expected number of $\mu$-DIS-induced $\pi^0$ produced in the final 
radiation length of the dump is 
\begin{equation}
    \Npot n_\mu^{(\mathrm{back})} A n_A \lambda_{\rm rl} \langle \sigma n_{\pi^0}\rangle
                           \sim 10^6
\label{eq:tertiary_pi0}
\end{equation}
where $ \langle \sigma n_{\pi^0}\rangle$ is given in~\cref{eq:DISCross}. The target (nucleon) density in iron is $A n_A \approx 4.7\times 10^{24}\;\mathrm{cm}^{-3}$. 

The number of kaons that are produced in a similar manner in the last kaon interaction length, and that decay beyond the dump is
\begin{equation}
  \begin{split}
    &\Npot n_\mu^{(\mathrm{back})} A n_A \lambda_{K\;\mathrm{int}} \langle \sigma n_{K_L}\rangle \mathrm{Br}(K_L\to 3\pi^0)\\
                            & \times \left[1 - e^{-(\zmax-\zdump)/(\gamma c\tau_{K_L} )}\right] \sim 10^4 
    \end{split}
    \label{eq:tertiary_KL}
\end{equation}
where the quantity the brackets is the probability of $K_L$ produced at the back of the dump 
to decay between $z = \zdump$ and $\zmax = 19\;\mathrm{m}$ (about 0.14 for the mean kaon energy produced in the DIS reaction). 

While the raw counts for the backgrounds in \cref{eq:tertiary_pi0,eq:tertiary_KL} appear problematic, it turns out that simple selections on photon energy and angle distributions can make these backgrounds significantly smaller. Nevertheless, as we will discuss in the next section, it will be challenging to completely suppress this source of background. 

Finally, we briefly mention a related background process: neutrino-induced DIS. In addition to muons, meson decays near the front of the dump also produce neutrinos, which are not attenuated or deflected by the FMAG. Like muons, these neutrinos can undergo DIS near the back of the dump, producing background diphotons. We found that the rate of such events is 5-6 orders of magnitude smaller than their muon-induced counterparts discussed above; the relative suppression arises because muon DIS is mediated by photon exchange, while neutrino DIS by the weak gauge bosons. As a result, neutrino-induced DIS contributes less than 10 ($<1$) $\pi^0$ ($K_L$) decays before applying any cuts, making this background sub-dominant to muon DIS.
\begin{figure}[t]
    \centering
    \includegraphics[width=0.47\textwidth]{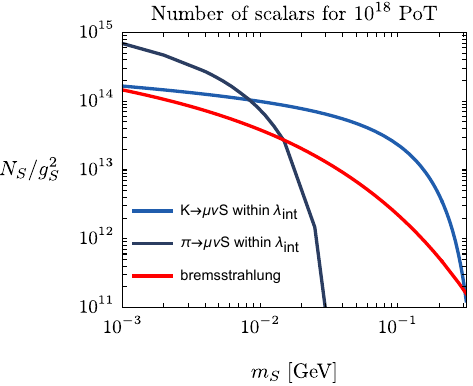} 
    \caption{Scalar yield as a function of the mass $m_S$ for $10^{18}$ POT from different production mechanisms without imposing any event selections.}%
    \label{fig:scalar_yield_channel_comparison}%
\end{figure}
\begin{figure*}[t]
    \centering
    \includegraphics[width=0.47\textwidth]{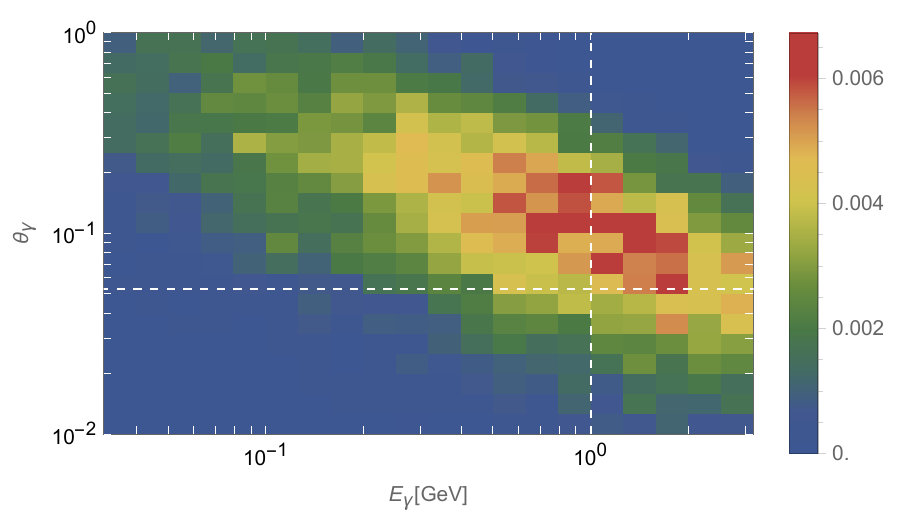}
    \includegraphics[width=0.47\textwidth]{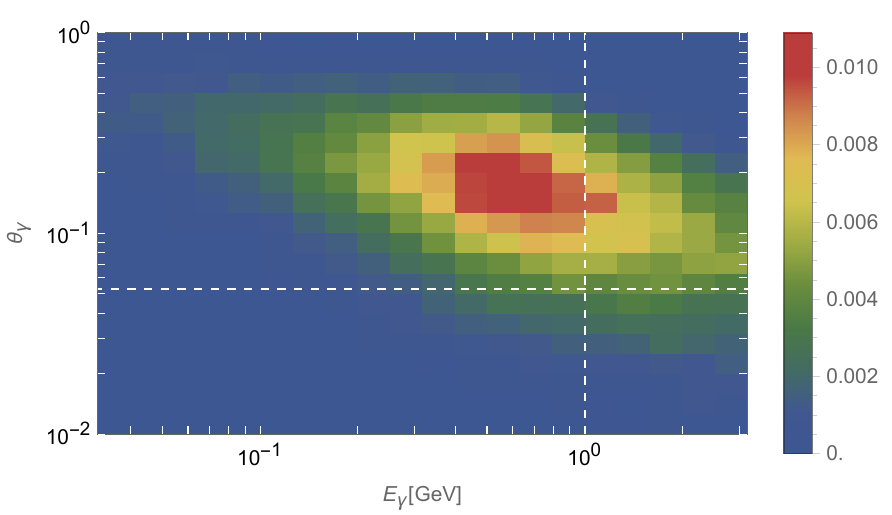}
    \caption{Histograms of signal photon energy and angle from $S\to\gamma\gamma$ with $S$ produced in rare kaon decays (left panel) and muon bremsstrahlung (right panel). $\theta_\gamma$ is the angle between the photon and the beam axis. In both panels $m_S = 100$ MeV and the bin normalization is such that they sum to 1. The vertical dashed white line corresponds to an energy cut of 1 GeV and the horizontal line corresponds to photons being roughly in geometric acceptance of the detector for scalars decaying just after the dump.}%
    \label{fig:signal_photon_energy_angle_distributions}%
\end{figure*}
\begin{figure}[t]
    \centering
    \includegraphics[width=0.47\textwidth]{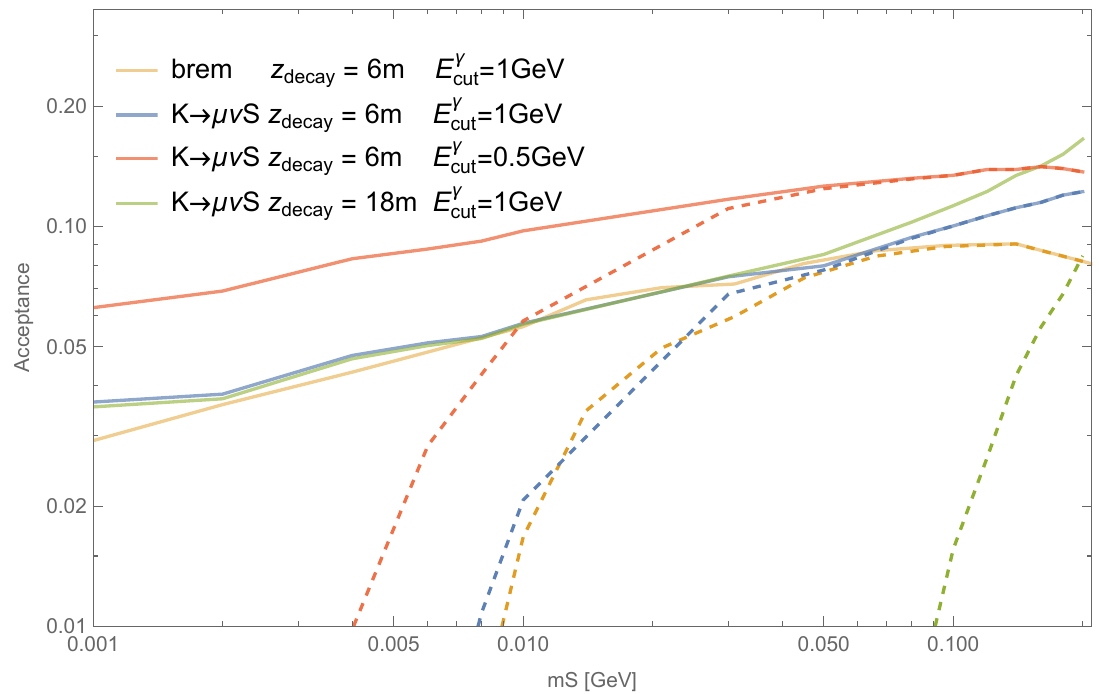}
    \caption{Acceptance probability of signal photons as a function of scalar mass for different $S$ production channels, decay positions and event selections. Dashed curves indicate a further $5.5$ cm separation cut was applied to the photons hitting the detector.}
    \label{fig:Kdecay_geometric_efficiency}%
\end{figure}

  \subsection{Background Uncertainties}
In the previous subsections we made several simplifying assumptions to make estimates of major background processes tractable: we did not consider the effects of electromagnetic (EM) or hadronic showers in the target, and we focused on production of mesons and muons in the first interaction length only. This means that the backgrounds are likely to be systematically underestimated. However, a careful treatment of these processes leading to higher meson and muon yields will also increase the signal rate. We therefore expect that signal over background rates should be similar to what we find here. In the remainder of this subsection we argue that our approximations are appropriate to a factor of a few in all of the processes considered above. 

First, let us consider the impact of including full EM and hadronic showers. Secondaries in EM showers can produce additional mesons and muons; however, the hadron photo- and electro-production cross-sections are at the microbarn level, much smaller than typical hadronic cross-sections of order of tens of millibarns. 
The cross-section for photon conversion into muons is at most $\sim 100$ $\mu$b at DarkQuest energies, which corresponds to a relative probability of $\sim 10^{-5}$ of converting into $\mu^+\mu^-$ compared to $e^+e^-$. This a much bigger penalty compared to restricting to muon production from $\pi^\pm$ decays in the first interaction length -- see \cref{tab:muons_photons_and_decays_in_first_interaction_length}.

Next, we estimate the impact of including hadronic showers by simulating protons incident on a 5 m iron target in \texttt{GEANT} and comparing the resulting pion yield to what we use. We find that the thick target produces about 4 times as many pions compared to our \pythia estimate. Minimum energy cuts bring the two estimates closer together.

Finally we comment on the meson ($K_L$) and muon (via $\pi^\pm$) production by the attenuated proton beam beyond the first interaction length. $K_L$'s produced deeper in FMAG have less material to pass through and therefore have a higher probability to decay in the signal region. However, the proton interaction length, $\lambda_{p\;\rm{int}}$, is slightly shorter than the meson interaction length, $\lambda_{K\;\rm{int}}$. This means that the $K_L$ flux is still exponentially attenuated, albeit with a significantly longer attenuation length $(1/\lambda_{p\;\rm{int}} - 1/\lambda_{K\;\rm{int}})^{-1} \sim 94\;\text{ cm}$. Therefore, the contribution of the dump beyond the first interaction length can be larger by a factor of $94 \text{ cm}/\lambda_{p\;\rm{int}} \sim 6$ compared to the estimate in~\cref{eq:KL_yield_estimate}.

Muon-induced backgrounds are also slightly underestimated since there will be both $\pi^\pm$ decays beyond the first interaction length and additional $\pi^\pm$ production by the attenuated proton beam. Similar arguments to the $K_L$ case above imply that allowing production and decay of $\pi^\pm$ anywhere in the dump would increase our estimate of the muon yield by a factor of $\sim 4$.

In summary, our background estimates, based on several simplifying assumptions, are expected to be accurate at the $\mathcal O(1)$ level, though likely systematically underestimated by a factor of a few. A more precise determination of the background rate is beyond the scope of this paper. However, a more accurate treatment of the relevant processes would also increase the signal yield by comparable factors. 
Therefore, we expect the reach presented here to be consistent with more precise determinations of background and signal rates.

\section{Results and Discussion}
\label{sec:results}

\begin{figure*}[t]
    \centering
\includegraphics[width=0.47\textwidth]{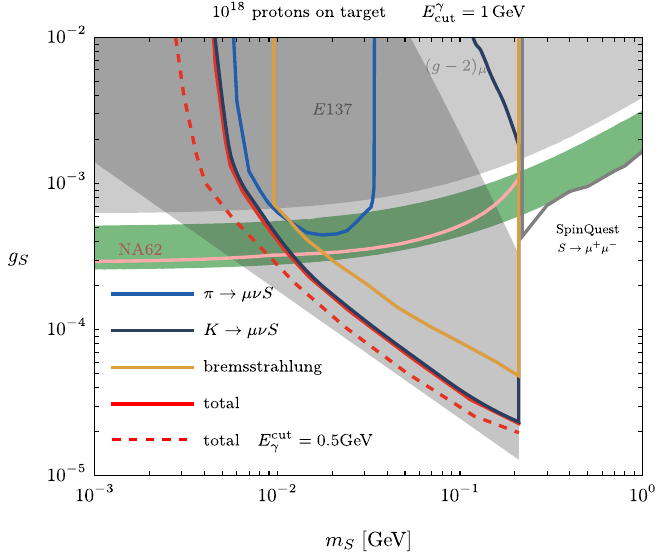}
\includegraphics[width=0.47\textwidth]{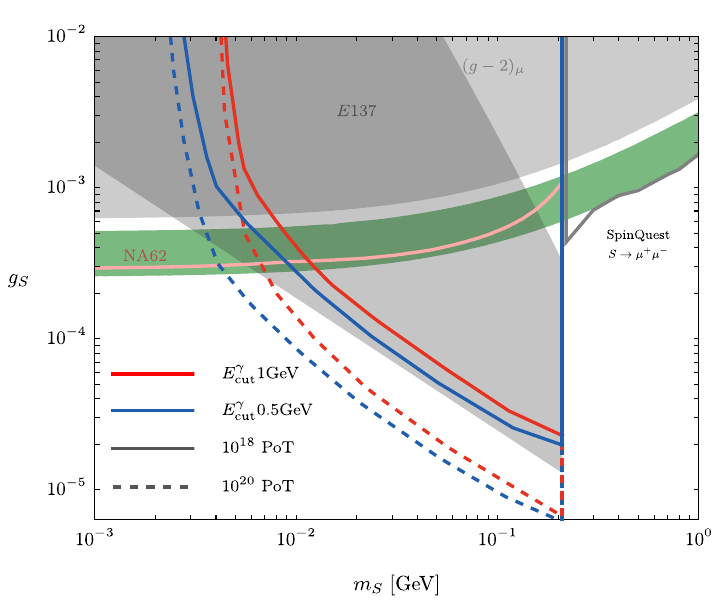}
\caption{Projected sensitivity of the DarkQuest experiment to the muon-philic scalar model in the $m_S-g_S$ plane for $m_S \leq 2m_\mu$. In the left panel we show contributions of different production channels to the sensitivity for $10^{18}$ POT (the dashed line is for $10^{20}$ POT). The signal photons are required to have $E_\gamma > 1\;\GeV$ and to be separated by more than 5.5 cm in the transverse direction when they enter the ECAL. In the right panel we show how the sensitivity changes with a lower cut on $E_\gamma$, and how the reach is improved with $10^{20}$ POT. In both plots the green band is the parameter space that can address the \gmtwo anomaly; the gray regions are excluded either by \gmtwo or by the recast of E137~\cite{Marsicano:2018vin}. We also show projections for the proposed resonant search at SpinQuest~\cite{Forbes:2022bvo} and at NA62~\cite{Krnjaic:2019rsv}.  
}%1
    \label{fig:reach}%
\end{figure*}
\begin{figure}[tbh]
    \centering
{\hspace{-0.6cm}{    \includegraphics[width=7.4cm]{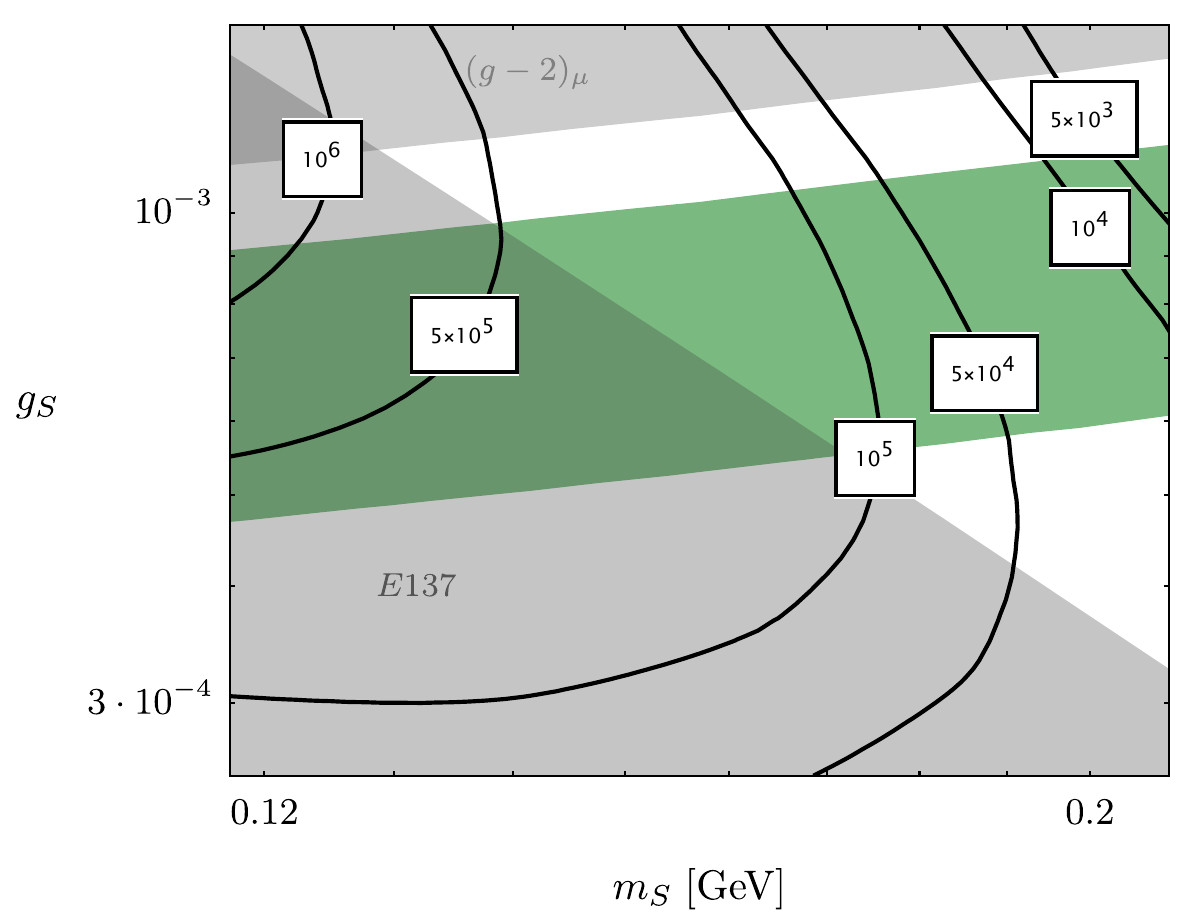}}}
        \includegraphics[width=7cm]{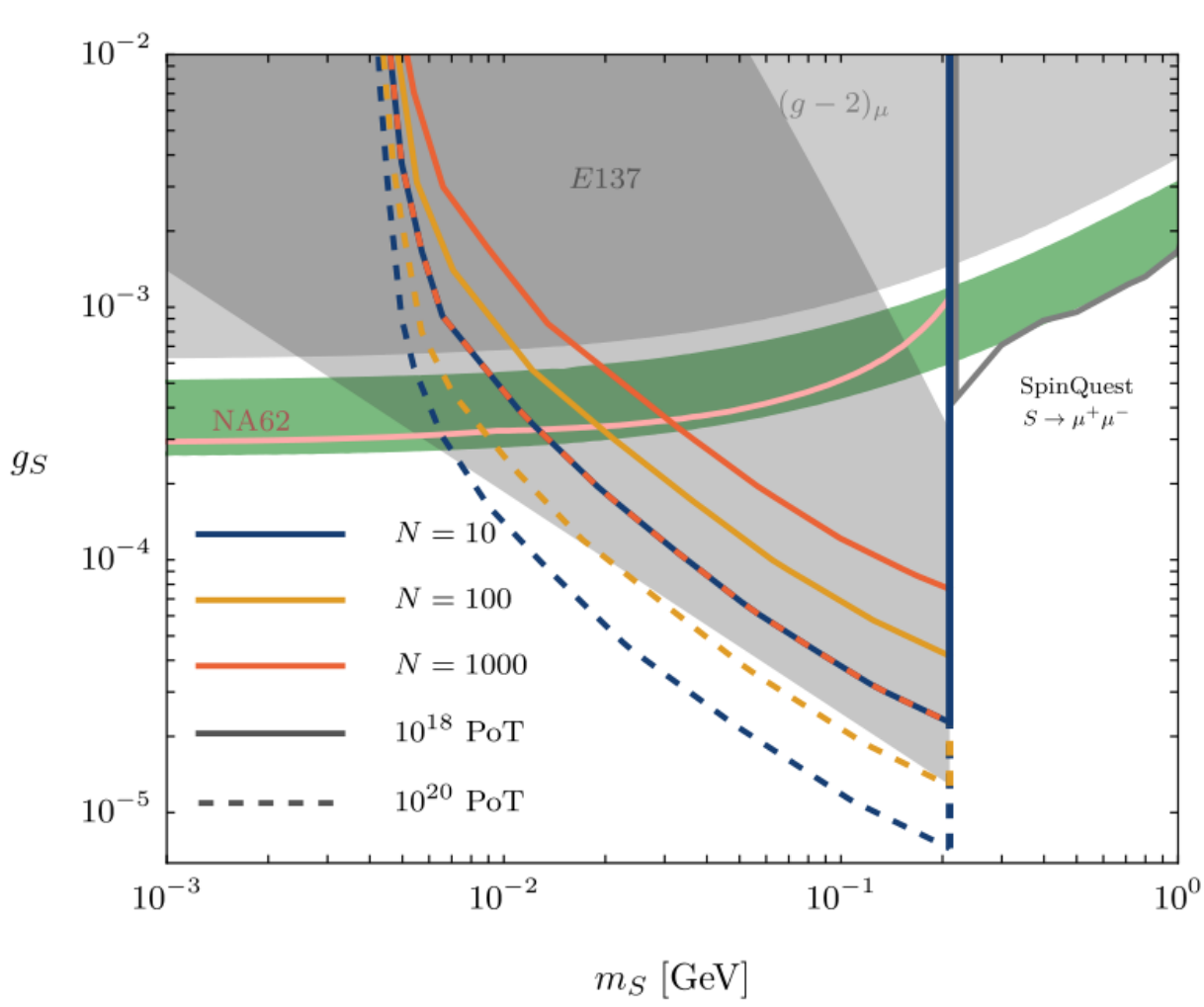}
    \caption{(Upper panel) Number of signal events for $10^{18}$ POT. (Lower panel) Contour lines for 10, 100, and 1000 events for both $10^{18}$ and $10^{20}$ POT. For both panels, we require two photons with $E_\gamma > 1\;\GeV$ each, separated by $\geq 5.5$ cm }
    \label{fig:event_yield_contours_g_minus_two_region}%
\end{figure}
\begin{figure*}[tbh]
    \centering
    \subfloat[\centering]{{\includegraphics[width=7cm]{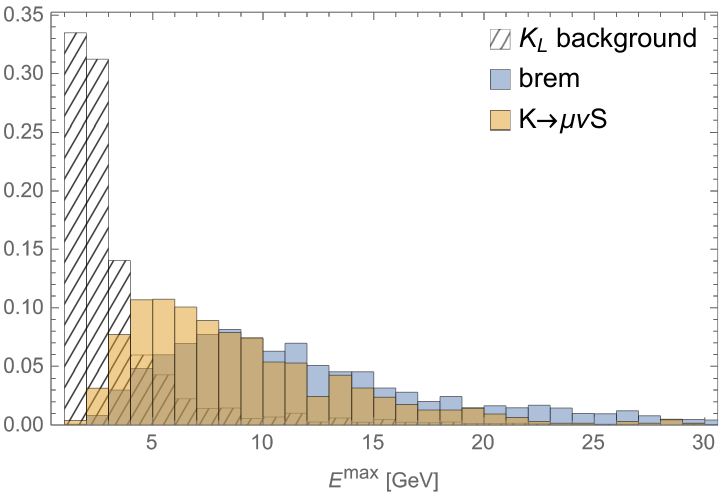} }}%
    \subfloat[\centering]{{\includegraphics[width=6.8cm]{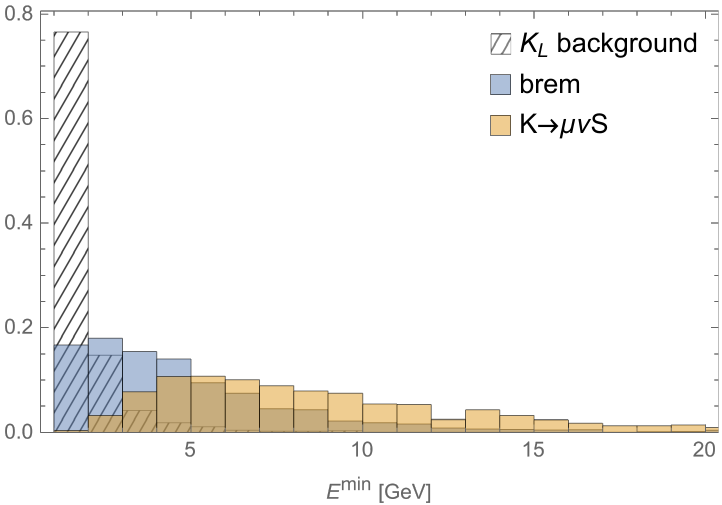} }}\\%
    \subfloat[\centering]{{\includegraphics[width=7cm]{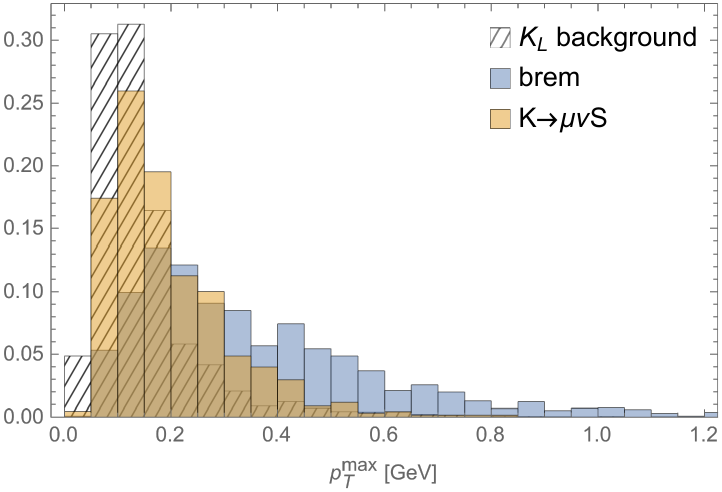} }}%
    \subfloat[\centering]{{\includegraphics[width=6.9cm]{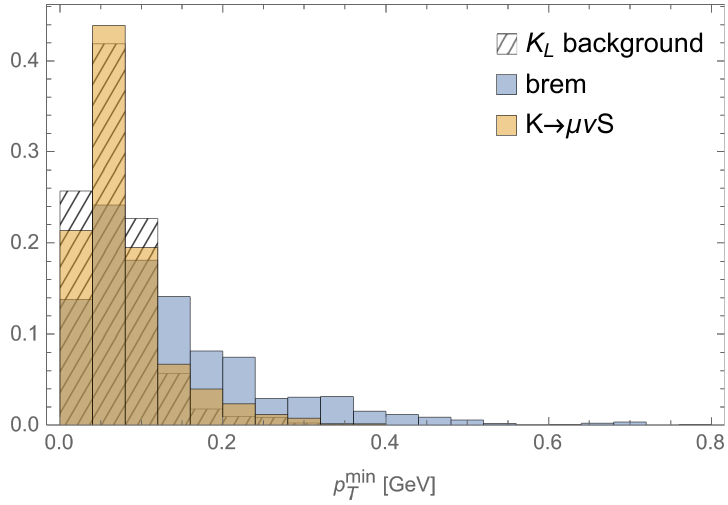} }}%

    \caption{Normalized histograms of kinematic variables for a parameter point at high scalar mass, inside the \gmtwo-favored region ($m_S=0.14\;\GeV, g_S\sim 10^{-3}$). The bremsstrahlung and the three body kaon decay $K\to\mu\nu S$ events are selected by requiring each photon in the pair have $E_\gamma > 1\;\GeV$ and $\geq 5.5$ cm separation in the ECal. The $K_L$ background events are selected by grouping the 6 photons into 5.5 cm square calorimeter towers and requiring that two distinct towers have a total $E_\gamma > 1\;\GeV$. The variables which are being plotted are then totaled for that tower. }
    \label{fig:signal_and_background_photon_distributions}%
\end{figure*}
\begin{figure}
    \includegraphics[width=0.48\textwidth]{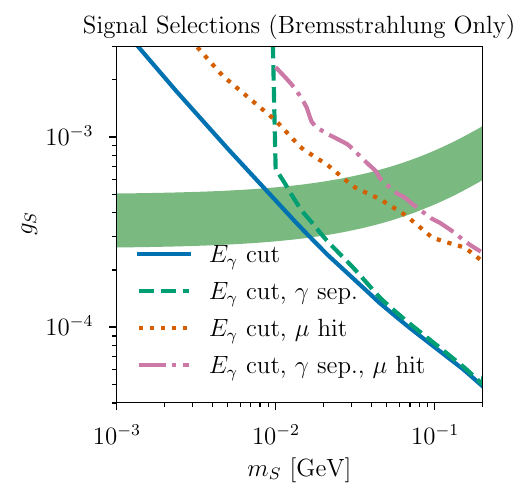}
    \caption{Comparison of the (background-free) experimental reach for different signal event selections for the bremsstrahlung production mechanism. All projections include a minimum photon energy cut of $1$ GeV; the green dashed (orange dotted) line also requires transverse photon separation of at least $5.5$ cm (a muon hit with $E_\mu > 1\;\GeV$ in the muon station). The pink dot-dashed line combines all of these selections. Even for this more restrictive set of cuts designed to eliminate the SM LLP background, DarkQuest maintains sensitivity to the \gmtwo band at $m_S \lesssim 2m_\mu$. \label{fig:brem_reach_vs_selections}}
\end{figure}
In this section we put together our signal and background simulations to project the sensitivity of future DarkQuest searches for muon-philic scalars 
in the $m_S  \lesssim 2m_\mu$ regime. First, we compare the raw $S$ yields from different production channels without making any selections in \cref{fig:scalar_yield_channel_comparison}. We observe that production from meson decays dominates the $S$ yield throughout our mass range. While rare $\pi^0$ decays produce the largest $S$ flux for $m_S \lesssim m_\pi - m_\mu$, we will see that even minimal event selections make this channel less important than the other two. Kaon decays produce more $S$ that $\mu$-bremsstrahlung despite the initial flux of $\mu$ being larger -- see \cref{tab:muons_photons_and_decays_in_first_interaction_length}. This occurs for two related reasons: first, the spectrum-averaged muon interaction probability in the dump is smaller than the $K^\pm$ branching fraction for the $m_S$ of interest; second, while the bremsstrahlung cross-section grows with initial muon energy, the probability of those muons to be produced in pion decays near the front of the dump is smaller, because of their higher boost. Thus the interaction probabilities are anti-correlated with the energy-differential muon flux, which is not captured in \cref{tab:muons_photons_and_decays_in_first_interaction_length}.

Examples of distributions of the energy and the angle between the photon and the beam axis are shown in \cref{fig:signal_photon_energy_angle_distributions} for the two dominant production mechanisms. The probability of these photons ending up in the detector acceptance depends on the precise location of the $S$ decay and any selections imposed on the events. In \cref{fig:Kdecay_geometric_efficiency} we show the effects of several simple cuts on energy and photon separation for a few fixed decay positions.  
We consider minimum photon energy cuts of 0.5 and 1 GeV as representative values, but looser selections are also possible~\cite{Apyan:2022tsd}. 
Requiring photon separation will enable the selection of events with exactly two photons which may be a useful handle for rejecting backgrounds such as $K_L \to 3\pi^0$.
The choice of 5.5 cm minimum transverse separation corresponds to the transverse granularity of the ECAL towers used by DarkQuest~\cite{PHENIX:2003fvo,Apyan:2022tsd}.
In \cref{fig:Kdecay_geometric_efficiency} we see that photon separation strongly reduces sensitivity to light scalars. As we discuss next this is not an issue since the novel parameter space that DarkQuest can test lies at larger masses. 
In \cref{fig:reach} we show background-free sensitivities for easy comparison with previous literature. The reach curves correspond to 10 signal event contours in the $m_S - g_S$ plane for phase 1 ($\Npot \approx 10^{18}$) and phase 2 ($\Npot \sim 10^{20}$). One remarkable feature of these projections is the sensitivity to relatively large couplings where the lab-frame lifetime of $S$ is expected to be significantly shorter than the FMAG depth of $5$ m. It arises due to the secondary muons' ability to interact deep in the dump, leading to $S$ production much closer to $z=5$ m as discussed in \cref{sec:signal_sim}. From the figure, we can see that already phase 1 could probe entirely the $(g-2)_\mu$ region at higher masses. Phase 2 will also have access to lower masses. We note that estimates of the DarkQuest sensitivity to the muonphilic scalar model have appeared in recent whitepapers~\cite{Harris:2022vnx,Gori:2022vri}. While the reach shown in \cite{Gori:2022vri} is in good agreement with the present results, \cite{Harris:2022vnx} made several simplifying assumptions (particularly limiting $S$ production to the front of the dump, and restricting $S$ decay to $500 < z < 900$ cm) which made the reach significantly weaker than what we find here.

The projections in~\cref{fig:reach} are valid only 
if the backgrounds discussed in \cref{sec:backgrounds} are reduced or eliminated. A simple, brute-force way to achieve this for the SM LLP backgrounds and diphotons from meson produced by the proton attenuated beam in the back of the dump is to include substantial extra shielding. Such a modification of the DarkQuest setup may not be practically feasible so it is worth considering other mitigation strategies. 

We note that while DarkQuest is sensitive to very small couplings, it does not surpass the (recasted) limits from E137~\cite{Marsicano:2018vin} in the long-lifetime regime for $N_p=10^{18}$ POT. In fact, the biggest impact of DarkQuest will be to probe the parameter space with $m_S\lesssim 2m_\mu$ that can explain \gmtwo; in this regime the couplings are substantial $g_S\lesssim 10^{-3}$, leading to signal rates of $\mathcal{O}(10^{3}-10^{6})$ as we show in the upper panel of \cref{fig:event_yield_contours_g_minus_two_region}. This suggests that the analysis can tolerate additional cuts that lower signal acceptance while still testing \gmtwo. It will be challenging for DarkQuest to access large new regions of parameter space at much lower masses and the precise reach will be affected by the exact amount of background. This is shown in the lower panel of \cref{fig:event_yield_contours_g_minus_two_region}, where we present the contour lines for 10, 100, and 1000 events for both $10^{18}$ and $10^{20}$ POT. We delineate a few background reduction strategies below. 
  \begin{table}
    \begin{tabular}{|L|L|L|L|L|L|L|L|}
      \hline
      N_\textrm{towers} & 0 & 1 & 2 & 3 & 4 & 5 & 6 \\
      \hline
      E^\textrm{tower}_\textrm{cut}=1.0\GeV & 55\% & 8\% & 8\% & 9\% & 10\% & 7\% & 3\% \\
      \hline
  \end{tabular}
  \caption{Distribution of the number of distinct calorimeter towers whose total energy deposition is above $E^\textrm{towers}_\textrm{cut}$ for photons produced from the $K_L\to 3\pi^0$ background. Similar numbers are obtained using the 0.5 GeV cut on the photons.}
  \label{tab:background-KL-pixel-efficiency}
  \end{table}
Requiring exactly two calorimeter towers with an energy deposition of at least 1 GeV eliminates a large fraction of the $K_L\to 3\pi^0$ background (see~\cref{tab:background-KL-pixel-efficiency} with a $92\%$ reduction). The remaining events have photon energy and $p_T$ distributions that are drastically different from those of the signal as we show in \cref{fig:signal_and_background_photon_distributions}. This suggests additional selections that can further reduce this background. For example by increasing the minimum photon energy cut to 2.8 GeV one can reduce the $K_L$ backgrounds by an order of magnitude. This selection can be further optimized by using different cuts for the lower and higher energy photons in the ECAL. For example, a lower energy photon cut of $E > 4$ GeV together with a higher energy photon cut of $E > 10$ GeV would reduce the background by a factor of 50 while reducing signal by a factor of 3.

Diphoton production in the last radiation length of the dump from the attenuated proton beam (\cref{sec:mesons_from_back_of_the_dump}) and displaced decays of $K_L$ (\cref{sec:sm_llps}) do not produce associated muons, while the dominant $S$ production channels do. Thus, selecting events with a muon can be an effective way of eliminating these backgrounds. In \cref{fig:brem_reach_vs_selections} we show how the background-free sensitivity changes with the addition of various cuts, including the muon selection.  While requiring a muon in the acceptance of the detector incurs a significant penalty on the signal rate, DarkQuest remains sensitive to the parameter space that can explain \gmtwo. The impact of the muon selection requirement was estimated by propagating each muon after it underwent bremsstrahlung in the FMAG through the remainder of the dump, through KMAG and to the muon station at $\sim 21$ m. The muon hit was counted if it had an energy over 1 GeV and it intersected the station 4 detectors~\cite{SeaQuest:2017kjt} (labelled ``muon station'' in \cref{fig:darkquest}).

Requiring a muon hit does not eliminate the muon DIS-induced neutral meson production at the back of the dump (see \cref{sec:mu_dis}). However, we find that simple geometric acceptance and photon energy selections reduce the estimated $\pi^0$ yield in \cref{eq:tertiary_pi0} by a factor of $> 200$. Similar conclusions hold for DIS-induced $K_L$ backgrounds: requiring exactly two photons with energy above a GeV in the geometric acceptance reduces this by a factor $> 1000$, even before cutting on the final state muon or photon separation. 

We have also explored a diphoton invariant mass cut to reduce backgrounds coming from $\pi^0$ decays. While excluding events with $m_{\gamma\gamma} \approx m_{\pi}$ does decrease these by a factor of at most $\sim 2$, the poor diphoton mass resolution discussed in \cref{sec:diphoton_mass_resolution} means that the signal yield is also negatively affected since DarkQuest's most important sensitivity lies in the window $m_\mu \lesssim m_S \lesssim 2m_\mu$, i.e., near the pion mass.

\section{Conclusions}
\label{sec:conclusions}

The proposed DarkQuest experiment offers a powerful, near-term test of scalars coupled to muons. In this paper, we studied di-photon signatures at DarkQuest and showed that DarkQuest will be able to probe new regions of parameter space that address the $(g-2)_\mu$ anomaly already with $10^{18}$ POT, for $m_S\leq 2 m_\mu$. 
The large number of secondary mesons and muons produced in the proton-dump collisions open multiple production channels for new light states whose displaced decays can be observed with the downstream detectors. We identified new production channels for scalars coupled to muons and background processes relevant for the displaced diphoton signatures at DarkQuest. We carefully modeled production and detection of signal and background events and outlined several strategies for achieving the projected sensitivities in \cref{fig:reach}.
Other searches (prompt $S\to\mu\mu)$ using SpinQuest will be able to probe higher masses~\cite{Forbes:2022bvo}. 
Several current and future experiments can also access similar \gmtwo parameter space. These include NA62~\cite{Rella:2022len}, NA64$\mu$~\cite{Gninenko:2018ter,Sieber:2021fue}, SHiP~\cite{Rella:2022len} and fixed-target-like analyses at the LHC multipurpose detectors~\cite{Galon:2019owl}. In particular, NA62 and NA64 are poised to do it this decade.\footnote{Recent results from NA64$\mu$~\cite{Andreev:2024lps} already exclude certain muonphilic vector models with invisible decays; it would be interesting to recast these results for the present model where the scalar decays outside of the detector, mimicking the invisible signal.} Compared to these experiments DarkQuest has sensitivity to large couplings which can close any gaps in the coverage of the \gmtwo-favored region, and it can identify the decay products of the muon-coupled force.

\section*{Acknowledgments}
We thank Cristina Mantilla Suarez, Christian Herwig, Yongbin Feng, Nhan Tran, Dean Robinson, Shirley Li, Wendy Taylor and Claudia Cornella for valuable discussions.
This research was enabled in part by support provided by the BC DRI Group, Compute Ontario and the Digital Research Alliance of Canada (\href{http://alliancecan.ca}{alliancecan.ca}). NB was supported in part by Natural Sciences and Engineering Research Council of Canada (NSERC). The research of SG and NH is supported in part by the U.S. Department of Energy grant number DE-SC0023093 and by the U.S. Department of Energy grant number DE-SC0010107. Part of this work was performed at the Aspen Center for Physics, which is supported by National Science Foundation grant PHY-1607611.

\onecolumngrid
\appendix
\section{Scalar-Photon Coupling}
\label{sec:scalar_photon_coupling}
In this section we calculate the form factor associated with the amplitude $S\to\gamma\gamma$ which gives rise to the photon coupling in~\cref{eq:photon_coupling}.
This amplitude provides the main $S$ decay channel (for $m_S < 2m_\mu$), and an $S$ production mechanism through secondary photons in 
the beam dump. The former involves two on-shell photons, while for the latter at least one photon is off-shell.
In general, this amplitude has both UV and IR contributions (from heavy matter that generates the defining $S\bar\mu\mu$ interaction and from the muon loop, respectively). 
The one loop amplitude with a single on-shell photon, generated by a generic fermion with mass $M$ and coupling $g_f$ to $S$ is 
\begin{equation}
  \mathcal{M}(S(p_S) \to\gamma(p_1)\gamma^*(p_2)) \equiv  \frac{ g_f \alpha Q^2}{2\pi M}  f_{1/2}\left(\frac{4M^2}{m_S^2}, \frac{p_2^2}{4M^2}\right)
   (p_1^\mu p_2^\nu - p_1\cdot p_2 \eta^{\mu\nu}),
\end{equation} 
where the form factor is
\begin{equation}
  f_{1/2}(\tau,\rho) =  \frac{\tau  \left((L_1^2-L_2^2) (\rho  \tau +\tau -1)+4 \rho  \tau  \left(L_1 \sqrt{\frac{1-\tau }{\tau ^2}} \tau -1\right)-4 L_2 \sqrt{(\rho -1) \rho } \tau +4\right)}{2 (\rho \tau -1)^2},
\end{equation}
with
\begin{equation}
\tau=\frac{4M^2}{p_S^2},~p_1^2=0,~\rho=\frac{p_2^2}{4M^2},
\end{equation}
and 
\begin{equation}
    L_1 = \ln \left(2 \sqrt{\frac{1-\tau }{\tau ^2}}-\frac{2}{\tau }+1\right),\;\;\; L_2=\ln \left(-2 \rho +2 \sqrt{(\rho -1) \rho }+1\right).
\end{equation}
This form factor reduces to the familiar expressions found for the fermionic contribution to $h\to \gamma \gamma$~\cite{Gunion:1989we}, when both photons are on-shell. 
For muons $g_f = g_S$ and $M = m_\mu$, but we can also use this result to estimate the contribution of heavy fermions $M^2 \gg m_S^2,\;p_2^2$ to the diphoton coupling ($g_{S\gamma\gamma}^{(\mathrm{UV})}$ in \cref{eq:photon_coupling_uv_and_ir}):
\begin{equation}
  \lim_{M\to \infty} f_{1/2}\left(\frac{4M^2}{m_S^2}, \frac{p_2^2}{4M^2}\right) = \frac{4}{3},
\end{equation}
so that 
\begin{equation}
  g_{S\gamma\gamma}^{(\mathrm{UV})} \sim \frac{2 g_f \alpha Q^2 }{3\pi M}.
  \label{eq:Sgammagamma_heavy_fermion}
\end{equation}
Note that generally $g_f \neq g_S$ (see next section) so the UV contribution to $S\gamma\gamma$ is not necessarily negligible compared to the muon one as we discuss in the next section.  

\section{An Ultraviolet Completion}
\label{sec:vll_uv_completion}
In this section we explore a simple ultraviolet completion for the effective operator in~\cref{eq:uv_completion1}, $- \frac{S}{M} H^\dagger L \mathbf{c}_S \ell_R^c,$ that generates 
the muon-philic interaction. This will allow to discuss how the low energy couplings are related to 
the underlying UV parameters, and to mention additional constraints on realistic scenarios. A similar scenario was considered in 
the Supplementary Materials of Ref.~\cite{Krnjaic:2019rsv}\footnote{For other scenarios, see e.g., \cite{De:2024tbo}.}.

The operator in~\cref{eq:uv_completion1} can be generated at tree-level by a heavy vector-like lepton (VLL) pair. 
The simplest possibility is a VLL pair of $SU(2)_L$ singlets $\psi$ and $\psi^c$, which have the same 
quantum numbers as the right-handed leptons $\ell^{c}_R$ in the SM (a completion with $SU(2)_L$ doublets is discussed in Ref.~\cite{Batell:2017kty}). 
The allowed renormalizable terms (in two-component notation) are  
\begin{equation}
\mathscr{L} \supset  -M \psi \psi^c - H^\dagger L^T y_\psi \psi^c  
 - y_S^\prime S \psi \psi^c - S \psi y_S^T \ell_R^c + \hc ,
\label{eq:vlm_lagrangian}
\end{equation}
where $L$ and $\ell_R^c$ are SM leptons (with flavor index suppressed), and the transposes are in flavor space. Since we consider a single generation of VLL, $M$ and $y_S^\prime$ are numbers, 
while $y_\psi$ and $y_S$ are vectors (with each entry corresponding to a SM fermion generation).
In the last two terms, we assumed that $\psi$ and $\psi^c$ have a Yukawa interaction with a scalar mediator $S$.\footnote{
The mass-mixing term $\mulpsi\ell_R^c\psi$ is allowed by gauge charges, but it can be rotated away by performing an orthogonal 
rotation on $(\ell_R^c\;\; \psi^c)$; this rotation results in redefinitions of the (unknown) parameters 
$M$, $y_S^\prime$, $y_S$, $y_\psi$ and $y_\ell$, which can then be relabelled.}
Integrating out $\psi$, and $\psi^c$ yields the effective operator 
\begin{equation}
  \mathcal{L} \supset \frac{S}{M} H^\dagger L^T y_\psi y_S^T \ell_R^c + \hc \to - \frac{v}{\sqrt{2}M} S L_L \mathbf{c}_S \ell_R^{c} + \hc ,
\end{equation}
where in the second step we transformed to the SM-lepton mass basis, which gives $\mathbf{c}_S \equiv - V_L^T y_\psi y^T_S V_R$ ($V_{L,R}$ are the 
left- and right-handed lepton unitary rotations that diagonalize their mass matrix). 
A muon-only interaction therefore requires that $\mathbf{c}_S  \propto \diag(0,1,0)$ as described in the introduction, necessitating a non-trivial 
flavor model. If this is realized then 
\begin{equation}
  g_S = \frac{y_S y_\psi v}{\sqrt{2}M}.
\end{equation}

There are several theoretical constraints on $g_S$ that follow from naturalness of the $S$ mass (i.e., the desire for quantum corrections 
to $m_S$ to not exceed $m_S$ itself). These can be derived either directly from the UV completion above or from the EFT interaction as in Ref.~\cite{Batell:2017kty}. 
The two most relevant quantum corrections to $m_S$ arise from 1) a two loop contribution from a Higgs-left handed lepton-right handed lepton loop; 2) from the one loop radiatively-generated $|H|^2 S^2$ interaction. Requiring that the two corrections do not induce a mass shift larger than $m_S$ leads to a theoretical upper bound on $g_S$:
\begin{equation}\label{eq:MBounndUV}
  g_S \lesssim \frac{4\pi v}{\sqrt{2}M} \min\left(\frac{\sqrt{2}m_S}{v},4\pi \frac{m_S}{M}\right),
\end{equation}
where the first (second) term comes from the two (one) loop correction. The one loop effect dominates for $M \lesssim 4\pi v /\sqrt 2 \simeq 2$ TeV. At the same time, $M$ cannot be too small in order to satisfy \cref{eq:MBounndUV} in the region of interest (see \cite{Batell:2017kty}). 

We can also estimate the UV contribution to the $S\gamma\gamma$ coupling, by using~\cref{eq:Sgammagamma_heavy_fermion}. 
At one loop, only the $y_S'$ coupling from \cref{eq:vlm_lagrangian} contributes, yielding
\begin{equation}
    g_{S\gamma\gamma}^{(\mathrm{UV})} \sim \frac{2 y_S' \alpha Q^2 }{3\pi M}.
\end{equation}
We see that $g_{S\gamma\gamma}^{(\mathrm{UV})}$ is not necessarily correlated with $g_S$, and it can 
therefore be comparable to $g_{S\gamma\gamma}^{(\mathrm{IR})}$ (generated by a muon loop) for 
$y_{S}' \gtrsim g_S (M/m_\mu)$.\footnote{This, however, would require a large hierarchy between dimensionless parameters.}

Finally let us briefly discuss high energy collider constraints on VLL. 
Recent theoretical studies of these models in the context of the LHC and future colliders include Refs.~\cite{Dermisek:2014qca,Kumar:2015tna,Bhattiprolu:2019vdu}.
The production rate (and the resulting bounds) depend strongly on the VLL $SU(2)_L$ representation~\cite{Kumar:2015tna}, with singlets usually 
being the least constrained. In the singlet case, the main production channel is $pp \to \gamma, Z^* \to \psi^+ \psi^-$.
The heavy leptons decay to SM particles via the interactions of~\cref{eq:vlm_lagrangian}.
Our model differs slightly from the minimal VLL scenarios studied in these works, 
as it features new production and decay mechanisms involving $S$. 
For example, in addition to the ``standard'' decays $\psi \to W \nu, \; Z \ell$, the VLL can decay $\psi \to S \ell$.  
Even more interestingly, interference of SM and $S$-mediated channels can significantly weaken existing constraints~\cite{Egana-Ugrinovic:2018roi}. 
The standard and $S$-mediated effects are parametrized by independent couplings, making it difficult to directly translate existing bounds on VLL models.
It would therefore be interesting to carry out a dedicated study for the VLL+$S$ model at the LHC. 

With these important caveats in mind, we note some constraints on the minimal VLL model (without $S$) that 
may be relevant for the full scenario: LEP ruled out many kinds of charged particles with mass below $\sim 100$ GeV~\cite{ADLO_SUSY,L3:2001xsz}. 
The best current LHC limits on the singlet VLL scenario are from ATLAS and CMS~\cite{ATLAS:2015qoy, CMS:2022nty}, with $M$ excluded in the ranges  $114-176$ GeV ($\psi \to Z+e/\mu$) or $125-150$ GeV ($\psi \to Z+ \tau$). 

\section{Muon Transport}
\label{sec:muon_transport}
In this Appendix, we describe our simulation of muon transport in the dump. We developed our own propagation code instead of using off-the shelf tools such as \texttt{GEANT} in order to have a simple way to incorporate rare processes relevant for signal and background events (e.g., $S$ bremsstrahlung and muon deep inelastic scattering). Simulations of these processes are performed using other codes (e.g., \texttt{MadGraph} and \pythia), so having minimal muon code enables a simple interface between them.

Muons produced near the front of the beam dump can propagate a significant distance before undergoing a dark bremsstrahlung reaction. 
The energy and angular distributions of the resulting muon-philic scalars are sensitive to the details of muon transport through the 
dump. Muon propagation is also relevant for predicting the rates of some of the backgrounds, such as DIS-induced meson production at the very back of the dump (see \cref{sec:backgrounds}).
Below we describe our treatment of muon production and propagation in the magnetic field, including energy loss and multiple scattering. 

We generate an initial muon sample using \pythia to model $pp$ collisions with incoming $120$ GeV proton beam, assuming the production takes place in the first nucleon interaction length. The primary proton flux is 
exponentially diminished beyond the first interaction length, and the surviving protons lose energy; neglecting these protons is therefore a conservative simplifying 
assumption.  

The muons come dominantly from charged pion decays. We require that the pions decay within the first 
pion interaction length, $\lambda_{\pi\text{ int}}$. The fraction of decayed pions grows linearly with distance (in the long life-time regime), 
while their flux is exponentially suppressed by $\exp(-z/\lambda_{\pi\text{ int}})$. As a result the total muon flux is well approximated by the 
first interaction length alone. This assumption allows us to neglect pion transport in the dump. 
The energy and transverse momentum distributions of pions and muons obtained this way are shown in~\cref{fig:secondary_distributions_from_120_GeV_pp}.

The secondary muons can undergo bremsstrahlung in the dump. Interactions close to the end of the dump are particularly interesting since they enable sensitivity to shorter scalar lifetimes. 
We propagate the muons through the FMAG in small steps, including the following effects:
\begin{itemize}
\item[a.] The 1.8 T magnetic field of the FMAG;
\item[b.] Ionization energy loss;
\item[c.] Multiple Coulomb scattering.
\end{itemize}
The magnetic field is included by solving the relativistic equation of motion in each step. 
Energy loss and multiple scattering are stochastic processes, so they are applied at the end of each step. We describe our treatment of these processes below. 

While $dE/dx$ is frequently used to estimate the energy losses of particles propagating through a medium, it is important to note that $dE/dx$ is an \emph{average} loss, appropriate for describing very large ensembles of particles. This average is sensitive to rare, high loss collisions. This does not mean that it is an accurate representation of the typical losses of individual particles~\cite{Workman:2022ynf}. Indeed, the definition of $dE/dx$ includes large losses due to rare high-energy transfer events, skewing the mean loss to larger values. For example, the most likely loss for a 5 GeV muon traversing 1 cm of iron is $10.6\;\MeV$, while $dE/dx$ is $14.6\;\MeV$~\cite{Workman:2022ynf}. In practice, this means that the losses of individual particles follow a distribution where the peak of the distribution (the most probable loss) is different from the average. In our case, energy loss fluctuations can be modeled by the Landau distribution (see Refs.~\cite{Talman:1978tc,Bak:1987cz,Bichsel:1988if,mcparland2016medical} for helpful reviews):
\beq
f(\Delta) = \frac{1}{2\pi i \xi} \int_{c-i\infty}^{c+i\infty} ds \exp{(\lambda s + s\ln s)},\;\;\; \lambda = \frac{\Delta - \Delta_p}{\xi} - 0.22278,
\label{eq:landau_dist}
\eeq
where $c$ is a positive number,  $\Delta$ is the energy loss, $\xi$ and $\Delta_p$ are given by~\cite{Workman:2022ynf}:
\begin{equation}
  \Delta_p = \xi \left[\ln \frac{2m_e \beta^2 \gamma^2}{I} + \ln \frac{\xi}{I} + 0.2 - \delta\right], 
  \;\; \xi = 0.153538\;\MeV\;\mathrm{cm^2}\;\mathrm{mol}^{-1} \frac{Z}{A} \frac{t}{\beta^2}, 
  \label{eq:landau_param}
\end{equation}
where $t$ is the thickness in g/cm$^2$ of the material traversed, $\beta$ the particle speed, and $A$ and $Z$ are the atomic mass and number, respectively. The material dependent parameters $I$ and $\delta$ are the mean excitation energy and density correction which are evaluated using the tabulated data of Ref.~\cite{Workman:2022ynf}.
The Landau distribution is an appropriate description of energy loss fluctuations when the losses are small compared to the maximum kinematically allowed loss and the initial kinetic energy ($\xi/E_{\mathrm{max}} \ll 1$ with $E_{\mathrm{max}}$ given in~\cite{Workman:2022ynf}), but large compared to typical atomic binding energies ($\xi/I \gg 1$)~\cite{mcparland2016medical}.  Both conditions are easily satisfied for muon energies of $\sim$ GeV and above, and for propagation distances of a few cm. \cref{eq:landau_dist} is heavy-tailed with an undefined mean, implying that particles are allowed to lose arbitrarily large amounts of energy. This is of course unphysical since there is a maximum energy loss dictated by kinematics. Thus, the distribution should be truncated, i.e., $\lambda \leq \lambda_{\mathrm{max}}$ for some $\lambda_{\mathrm{max}}$. We implement such a truncation by requiring that the mean energy loss of the truncated distribution is equal to $dE/dx$. Sampling from this truncated distribution is achieved using a modified version of the \href{https://github.com/SengerM/landaupy}{landaupy} package. We checked that our implementation of energy losses reproduces those of \texttt{GEANT} for muons with initial energies from 1 to 100 GeV propagating through iron targets of 1 to 500 cm, covering the parameter ranges relevant for DarkQuest.  

Multiple scattering is modeled by applying a transverse momentum kick drawn from a Gaussian approximation to the Moliere distribution of scattering angles. The variance $\sigma^2$ of this distribution is given in Sec. 7 of Ref.~\cite{Lynch:1990sq}:
\begin{equation}
  \sigma^2 = \frac{\chi_c^2}{1+F}\left[ \frac{1+v}{v} \log(1+v)-1\right],\;\; \chi_c^2=  \frac{4\pi n_A t \alpha^2 Z(Z+1)}{(p\beta)^2}  , \;\;    v = \frac{\Omega}{2(1-F)}, \;\; F=0.98 
\end{equation}
where $n_A$ is the atomic number density, $p$ is the particle momentum and $\Omega$ is the mean number of scatters
\begin{equation}
    \Omega = \frac{7800~ t ~(Z+1) Z^{1/3} }{\beta^2 A \left[1+ 3.34 \left(\frac{Z \alpha}{\beta}\right)^2 \right]}
\end{equation}
with $\beta$, $t$, $Z$, $A$ defined above.
The Gaussian approximation provides an adequate description of multiple Coulomb scattering when $\Omega \gg 1$; this is the case for $t \gtrsim 10^{-3} X_0$, $\beta=1$ and arbitrary $Z$, where $X_0$ is the radiation length~\cite{Lynch:1990sq}.

We chose a 1 cm step size to ensure the validity of the various approximations for the treatments of ionization and multiple scattering. 
Specifically, 1 cm is large enough such that the number of Coulomb collisions is large and the Gaussian approximation is valid, but small enough that the losses do not significantly affect the muon trajectory in a single step. A larger step size is also desirable for computational speed. We 
have verified that this algorithm reproduces the \texttt{GEANT4}-transported muon spectrum used in Ref.~\cite{Forbes:2022bvo}.\footnote{We thank Cristina Mantilla Suarez and Christian Herwig for extensive discussions regarding this check.} It also reproduces the spectra of mono-energetic muons propagated through $5$ m of iron (this cross-check eliminates the convolution of propagation effects with the primary muon spectrum). Note that while \cite{Forbes:2022bvo} focused on a dimuon resonance search with a mono energetic muon beam, here we take into account the full muon spectrum from beam-target interactions and its propagation through the dump. 

The muons are thus transported until they either leave the dump or reach an energy below a GeV 
(in our reach estimates we always require the signal photons to each have an energy of at least 500 MeV, which implies a minimum muon energy of $\mathcal{O}(1)$ GeV).
The magnetic field and energy loss are by far the most important effects: The $p_T$ imparted by the $B$ field is $\sim 2.6\;\GeV$ for relativistic muons that traverse the entire FMAG; in practice the $p_T$ fluctuates around this value due to multiple scattering and energy losses; it can also be substantially smaller if the muon ranges out inside the dump. The most probable energy loss ($\Delta_p$ in \cref{eq:landau_param}) is $\sim 7\;\GeV$ after traversing the FMAG. 

\begin{figure}
  \begin{center}
    \includegraphics[width=0.95\textwidth]{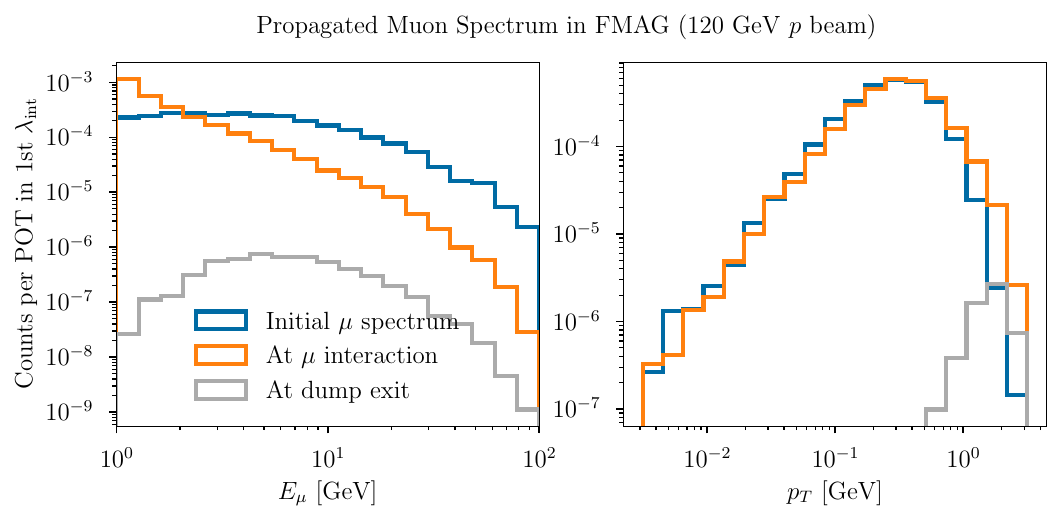}
  \end{center}
  \caption{Muon spectra at creation (blue lines), at the $\mu N \to \mu N + S$ interaction (orange lines), and at FMAG exit (gray lines). The initial spectra are generated using \pythia (see~\cref{fig:secondary_distributions_from_120_GeV_pp}) and transported as described in the text. The transport includes the effects of energy loss, multiple scattering, and the magnetic field which are evident in the left and right panels, respectively. Note that blue and orange histograms sum to $0.0028$ per POT, which is smaller than the total muon yield of $0.01$ per POT in \cref{tab:muons_photons_and_decays_in_first_interaction_length} and the corresponding histogram in \cref{fig:secondary_distributions_from_120_GeV_pp}; this is because here we impose $E_\mu > 1 \;\GeV$. The gray histogram contains a smaller number of muons still as not all muons are able to penetrate the entire FMAG. 
  \label{fig:muon_before_and_after_interaction}}
\end{figure}

The phase-space trajectories obtained using the above procedure are used to sample the momenta and interaction vertices at the interaction time. 
The probability distribution of the muon path length $\ell$ through the FMAG at interaction is given by (in the small coupling limit)
\beq
\frac{dP}{d\ell} \approx n_A \sigma_{\rm{Brem.}}(E_\mu(\ell)),
\eeq
where $n_A$ is the number density of target nuclei, $\sigma_{\mathrm{Brem.}}$ is the $\mu N \to \mu N + S$ cross-section, and $E_\mu(\ell)$ is obtained from the muon trajectory. 
This distribution is sampled using the inverse transform method. In~\cref{fig:muon_before_and_after_interaction} we show 
the energy and transverse momentum spectra of the initial muons, the muons at their interaction positions and the muons that traverse the entire FMAG without interacting. We clearly see the 
effects of transport, as the muons typically have a lower energy and larger $p_T$ the more material they traverse. In our simulation, we force 
every generated muon to interact within the FMAG. In this approach, each event must be weighted by the probability of the muon 
do so:
\begin{equation}
  n_A \int_{\mathrm{FMAG}} \sigma_{\rm{Brem.}}(E_\mu(\ell)) d\ell. 
\end{equation}
The geometric acceptance-weighted sum of these probabilities (normalized to the total number of simulated $pp$ collisions) gives the 
total $S$ yield (see \cref{eq:NsignalMuon}).

\section{Rare Meson Decays}
\label{sec:meson_decay_details}

The full result for the three-body decay of the kaon to produce a scalar is given by,\footnote{ The Mathematica version of this expression can be found \href{https://gist.github.com/nickhamer/ade3cfecc26a1b63e75e4df60d780097}{here}.}
\begin{align}
    &\Gamma\left( K \rightarrow \mu \nu S \right) = \frac{1}{(2\pi)^3} \frac{(G_f f_K V_\textrm{us})^2}{4 m_K} g_S^2 \quad \Bigg\{ \Bigg. \nonumber \\
    &\frac{1}{48 m_K^2} \left(
        2m_K^4 + 26 m_K^2 m_S^2+ 2 m_S^4 - m_\mu^2 \left(115 m_K^2 + 61 m_S^2 \right) + 119 m_\mu^4
    \right) \lambda^{1/2} \left(m_K^2, m_\mu^2, m_S^2 \right) - \nonumber \\
    &\frac{1}{8 m_K^2 m_\mu^2} m_S^2 \left( m_K^2 - m_\mu^2 \right) \left(
        2 m_K^2 m_\mu^2 + m_K^2 m_S^2 - 18 m_\mu^4 + 3 m_\mu^2 m_S^2
    \right)  g \left( m_\mu^2/m_S^2 \right)  \tan^{-1}\left[\frac{ g \left( m_\mu^2/m_S^2 \right) \, \lambda^{1/2} \left(m_K^2, m_\mu^2, m_S^2 \right)}{m_K^2 - m_S^2 + 3 m_\mu^2} \right] - \nonumber \\
    &\frac{1}{8 m_\mu^2} m_K^2 \left( m_S^2 - m_\mu^2 \right)^2 \log \left(
        \frac{m_K^2 \left( m_S^2 + m_\mu^2 \right) + \left( m_S^2 - m_\mu^2 \right) \left[ -m_S^2 + m_\mu^2 + \lambda^{1/2} \left(m_K^2, m_\mu^2, m_S^2 \right) \right]}
        {m_K^2 \left[ m_K^2 - m_S^2 - m_\mu^2 - \lambda^{1/2} \left(m_K^2, m_\mu^2, m_S^2 \right) \right] }
    \right) + \nonumber \\ \nonumber
    &\frac{1}{8 m_K^2 m_\mu^2}
        \Big[ \Big. \qquad  m_K^4 \left( m_S^4 - 4 m_\mu^2 m_S^2 + 8 m_\mu^4 \right) 
            - \, m_K^2 m_\mu^2 \left( 2 m_S^4 -24 m_S^2 m_\mu^2 + 16 m_\mu^4 \right) 
            + \, m_\mu^4 \left( 3 m_S^4 - 36 m_\mu^2 m_S^2 + 8 m_\mu^4 \right)
    \quad \Big. \Big]\times \\
    &\log \left( \frac{2 m_S m_\mu}{m_K^2 - m_S^2 - m_\mu^2 - \lambda^{1/2} \left(m_K^2, m_\mu^2, m_S^2 \right)} \right) \Bigg. \Bigg\},
\end{align}

where
\begin{align}
    g \left( m_\mu^2/m_S^2 \right) &\equiv \sqrt{ 4 m_\mu^2/m_S^2 - 1 }, \\
    \lambda^{1/2} \left(m_K^2, m_\mu^2, m_S^2 \right) &\equiv \sqrt{m_K^4 + m_S^4 + m_\mu^4 - 2m_K^2 m_S^2 - 2m_K^2 m_\mu^2 - 2 m_S^2 m_\mu^2},
\end{align}
where $f_K=156\;\MeV$ is the kaon decay constant. A similar expression holds for the corresponding pion three-body decay.

%%%%%%%%%%%%%%%%%%%%%%%%%%%%%%%%%

%\bibliographystyle{JHEP}
\bibliography{refs}

\end{document}